\documentclass[prl,superscriptaddress,twocolumn,amsmath,nofootinbib]{revtex4-2}
\usepackage{epsfig}
\usepackage{xcolor}
\usepackage{bm}
\usepackage{amsmath}
\usepackage{amsfonts}
\usepackage{dsfont}
\usepackage{placeins}
\usepackage{tikz}
\usetikzlibrary{positioning,arrows.meta}

\tikzset{
    Bnode/.style={draw,rounded corners,minimum width=3.1cm,
        minimum height=0.85cm,align=center},
    rel/.style={-{Latex[length=2mm]}, thick}
}

\begin{document}

\title{Linear Breit-Wheeler pair production in the search for axion-like particles }
\author{T. Rook}
\affiliation{Department of Physics, University of Oxford, Clarendon Laboratory, Parks Road, Oxford OX1 3PU, United Kingdom}
\author{W. Wu}
\affiliation{Department of Physics, Imperial College London, London SW7 2AZ, United Kingdom}
\author{S. P. D. Mangles}
\affiliation{Department of Physics, Imperial College London, London SW7 2AZ, United Kingdom}
\author{S. J. Rose}
\affiliation{Department of Physics, University of Oxford, Clarendon Laboratory, Parks Road, Oxford OX1 3PU, United Kingdom}
\affiliation{Department of Physics, Imperial College London, London SW7 2AZ, United Kingdom}
\date{\today}

\begin{abstract}
Assuming the existence of axion-like particles (ALPs), the standard Quantum Electrodynamics (QED) amplitudes for the linear Breit-Wheeler (BW) pair production process are shown to be supplemented by an additional ALP-mediated amplitude. We compute the coherently combined cross section and show that, for pseudoscalar ALPs, interference with QED produces an asymmetric Fano-type resonance.   
We propose experimental observation of the BW process with real photons as a complementary probe of MeV-scale ALPs, coupling jointly with photons and electron-positron pairs, on the basis that it occupies a kinematic region not readily accessible to other ALP search methods.

\end{abstract}

\maketitle
Since its inception in the 1920s \cite{Dirac1927}, and its subsequent development \cite{Tomonga1946,Dyson1949,Schwinger1948,Feynman1949}, Quantum Electrodynamics (QED) has been extraordinarily successful and is widely regarded  as one of the most rigorously tested theories in physics, accounting for the Lamb shift in atoms \cite{Lamb1947,Bethe1947} and providing ultra-high precision benchmarks for measurements of the anomalous magnetic moment of the electron \cite{Kusch1948,Aoyama2012,Hanneke2008}. 

Despite this success, some of the simplest predictions of QED have historically been challenging to experimentally isolate. A striking example is the direct conversion of two photons into an electron–positron pair,
\begin{equation*}
    \gamma + \gamma \rightarrow e^- + e^+.
\end{equation*}
This was first predicted in the 1930s by Breit and Wheeler \cite{Breit1934} who calculated the cross section, but lamented that it was ``\textit{hopeless to
try to observe the pair formation in laboratory
experiments}". Since then, the Breit-Wheeler (BW) process has only been observed indirectly (with virtual or quasi-real photons), through experiments involving the high-energy collisions of heavy ions \cite{STAR2021,ATLAS2017,ATLAS2021,CMS2025} and in an astrophysical context is understood to be responsible for the opacity of the universe to high energy photons \cite{Nikishov1962,Gould1967}.

The kinematics of the process require that the center of momentum (COM) energy be in excess of twice the electron mass, $\sqrt{s}\geq2m_e$ where $s = 4E^2$ and $E$ is the photon energy in the COM frame, working in natural units where $\hbar=c=1$. Direct observation is suppressed by the weakness of the electromagnetic coupling and by the practical difficulty of producing two sufficiently intense, collimated photon beams in the MeV regime. However, recent advances have made these difficulties surmountable \cite{Pike2014,Kettle2021,Watt2025}. 

In the coming years, we expect further experimental tests of the validity of QED through observations of the BW process with real photons \cite{Pike2014,Kettle2021,Watt2025, Esnault2021}, as well as continued interest in the BW process with virtual photons, stimulated through ultra-peripheral collisions. Calculations of the QED contribution have been made in  various settings \cite{Breit1934,Song2025,Voisin2017,Cabral2023}, however, further understanding of the experimental signatures of potential beyond-the-Standard-Model (BSM) physics which might appear during photon collisions is crucial, for the proper design and analysis of experiments. Examples of this are dark photons \cite{Wong2021}, millicharged fermions \cite{Davidson2000}, Born-Infeld theory \cite{Davila2014} and the presence of hypothetical charge-neutral, massive pseudoscalar (or scalar) axion-like particles (ALP). 

\begin{figure*}[]
    \centering
    \includegraphics[width=0.99\linewidth]{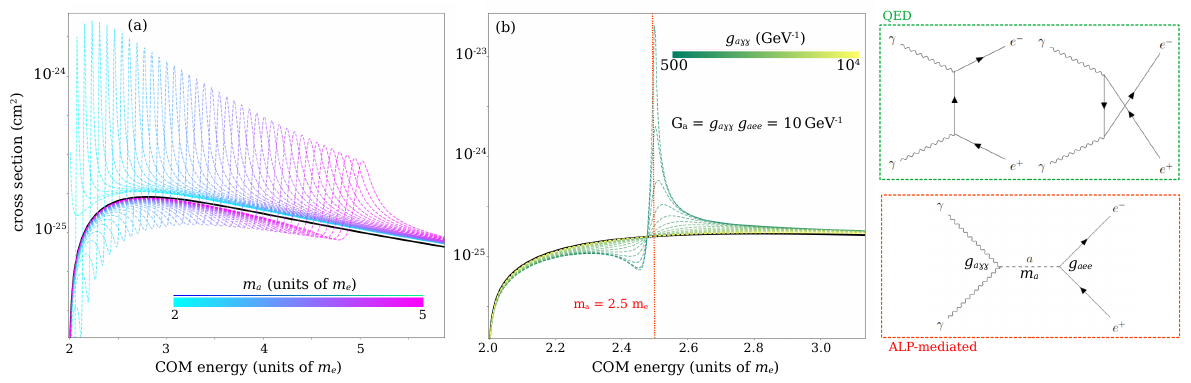}
    \caption{(a) and (b): QED Breit-Wheeler cross section (solid black line) alongside the combined ALP-mediated and QED cross section (dashed colored lines) with (a) ALP mass varied while $g_{aee}=1\times10^{-2}$ and $g_{a\gamma\gamma}=1\times10^{3}$~GeV$^{-1}$ and (b) ALP-photon coupling indicated by the colorbar (both $g_{a\gamma\gamma}$ and $g_{aee}$ are varied while their product is held constant) and $m_a=2.5m_e$. The red vertical dashed line in (b) indicates the ALP mass and we can see that a resonance appears in the cross section at this COM energy. The three Feynman diagrams for the process, used to calculate the cross section are shown on the right hand side of the figure.}
    \label{fig:xsection}
\end{figure*}

The axion was originally suggested as a BSM solution to the ``strong CP problem" \cite{Peccei1977a,Peccei1977b,Peccei2008,Kim2010}, with some ALPs being potential candidates for dark matter \cite{Preskill1983}. Naturally, a vast theoretical and experimental effort has been devoted to searching, in astrophysics \cite{Sikivie1983,Galanti2022} and in the lab \cite{Redondo2011},  for these particles and constraining the mass $m_a$, and strength of coupling with the quantum fields associated with Standard-Model particles \cite{AxionLimits}, in parameter space. Here, we focus on QED, so the relevant effective coupling parameters are the ALP-photon coupling $g_{a\gamma\gamma}$ and the ALP-electron coupling $g_{aee}$, which may take positive or negative values. These appear in the pseudoscalar ALP-QED Lagrangian,
\begin{equation}\label{eq:Lagrange}
    \begin{split}
    \mathcal{L}(a,\psi,A) =&  \mathcal{L}_{\mathrm{QED}} + \frac{1}{2}\partial_\mu a \partial^\mu a - \frac{1}{2}m_a^2a^2 \\
     -& \frac{1}{4}g_{a\gamma\gamma} a F_{\mu\nu}\tilde{F}^{\mu\nu} - i g_{aee}  a \bar{\psi}\gamma_5\psi,
\end{split}
\end{equation}
where $F_{\mu\nu} = \partial_\mu A_\nu - \partial_\nu A_\mu$,  $\mathcal{L}_{\mathrm{QED}}$ is the QED Lagrangian,
\begin{equation}
    \mathcal{L}_{\mathrm{QED}}(\psi,A) = -\frac{1}{4}F_{\mu\nu}F^{\mu\nu} + \bar{\psi}(i\gamma^\mu D_\mu - m_e)\psi,
\end{equation}
and $a$, $\psi$ and $A$ are the ALP, electron and photon field, respectively. In particular, this is the Lagrangian of a low-energy effective field theory (EFT) for a pseudoscalar ALP which is restricted to coupling with the electron, positron and photon fields only. Here, we can see that the ALP interacts with QED by coupling with pairs of photons and with electron-positron pairs. Therefore, the t- and u-channel QED processes, represented by the Feynman diagrams in the green dotted rectangle on the right-hand side of Fig.~\ref{fig:xsection}, are supplemented by an 
\FloatBarrier
\noindent ALP-mediated s-channel process (red dotted rectangle on the right-hand side of Fig.~\ref{fig:xsection}).

From here, we calculate the pseudoscalar-ALP-mediated BW cross section (derivation in Appendix A),
\begin{equation}\label{eq:axion_crossSection}
    \sigma_a = \frac{(g_{a\gamma\gamma}g_{aee})^2}{64\pi}\frac{ s^2\beta}{|s - m_a^2 +i\sqrt{s}\Gamma(s)|^2},
\end{equation}
where $\beta = \sqrt{1-4m_e^2/s}$ and
which, at first glance, appears to only depend upon the product of the different coupling parameters. 

However, there is a more complicated dependence which enters through the decay rate $\Gamma$ \cite{Dent2020} present in the denominator as a running width obtained by evaluating the imaginary part of the ALP self-energy,
\begin{equation}\label{eq:lifetime}
\begin{split}
    \Gamma(q^2) &= \Gamma_{a\rightarrow\gamma\gamma}(q^2)+\Gamma_{a\rightarrow e^+e^-}(q^2)\\
    &= \frac{g_{a\gamma\gamma}^2q^3}{64\pi} + \frac{g_{aee}^2q}{8\pi}\sqrt{1-\frac{4m_e^2}{q^2}}\Theta(q^2-4m_e^2),
\end{split}
\end{equation}
which determines the width of the resonance. 
The comparison of $\sigma_a$ with the standard QED cross section \cite{Jauch1976,Song2025},
\begin{equation}\label{eq:qed_crossSection}
    \sigma_{QED} = \frac{\pi r_e^2}{2}(1-\beta^2)\left[2\beta^3-4\beta+(3-\beta^4)\ln{(\frac{1+\beta}{1-\beta})}\right],
\end{equation}
where $r_e=\alpha/m_e$ is the classical electron radius and $\alpha = e^2/4\pi$ is the fine structure constant, gives a rough idea of the importance of the ALP contribution for different parameters and COM energy. A complete comparison, however, depends upon how the quantum mechanical amplitudes combine coherently, for the different pathways, 
\begin{equation}\label{eq:coherent}
    \frac{d\sigma}{d\Omega} = \frac{\beta}{256\pi^2s} 
    \sum_{i,j,k,l}
    \lvert \mathcal{M}^{(ijkl)}_{QED}(s,\theta,\phi)+ \mathcal{M}^{(ijkl)}_{a}(s) \rvert^2 .
\end{equation}
 Only this way can one determine whether the amplitudes interfere constructively or destructively (enhancing or suppressing $\sigma_\mathrm{QED}$, respectively). In particular, the typical Lorentzian profile of the Breit-Wigner resonance \cite{Breit1936} is supplanted by an asymmetric Fano-type profile \cite{Fano1961} caused by the interference between the resonant axion-mediated amplitude and the non-resonant ``continuum" QED amplitude and 
\begin{equation}
    \sigma_\mathrm{int} = \frac{1}{4\pi} \frac{ e^2 g_{a\gamma\gamma}g_{aee}m_e  (s-m_a^2)}{(s-m_a^2)^2+s\Gamma(s)^2}\tanh^{-1}(\beta).
\end{equation}
is the interference contribution to the total cross section, $\sigma=\sigma_\mathrm{QED}+\sigma_a+\sigma_\mathrm{int}$, with a detailed discussion of the calculation provided in Appendix C. Contrasting cross sections resulting from the presence of a scalar particle are provided in Appendix D.

The effect of including an ALP in the cross section is presented in Fig.~\ref{fig:xsection} for a range of ALP parameters (a more detailed breakdown of the impact of the interference term is provided by Fig.~\ref{fig:interference} in Appendix C). For $\vert G_a\vert = \vert g_{a\gamma\gamma}g_{aee}\vert$ less than about $10$ GeV$^{-1}$, the only significant deviation of $\sigma$ from $\sigma_\mathrm{QED}$ occurs around COM energy $m_a$. In  Fig.~\ref{fig:xsection}(a) the Fano-profile resonance is shown to move to increasing COM energy with increasing $m_a$, while the degree of enhancement subsides as $m_a$ reaches higher energies. Fig.~\ref{fig:xsection}(b) shows that for a fixed $m_a$ and a fixed $G_a$, the degree of resonance depends on the precise combination of $g_{a\gamma\gamma}$ and $g_{aee}$. In the regime considered here, the rate $\Gamma$ is particularly sensitive to $g_{a\gamma\gamma}$, so the resonance is enhanced when $g_{a\gamma\gamma}$ is decreased. 

The total cross section (integrated over solid angle) as presented in Fig.~\ref{fig:xsection} is precisely the quantity which is relevant in a heavily asymmetric beam setting, without polarization control (see experimental setup proposed in refs.~\cite{Pike2014,Kettle2021}). Here, the COM frame will be Lorentz boosted into the lab frame such that the pair produced electrons and positrons are collimated within a narrow cone around the high-energy photon beams axis. The angle resolved differential cross section (DCS) in the COM frame is less relevant here. Alternatively, collisions between beams which are symmetric in photon energy will lead to scattering at a range of angles determined by the angle resolved differential cross sections. Notably, both ALP DCS are totally isotropic while the QED DCS exhibits significant angular variation. The polarization dependence of the ALP DCS is provided in Appendix B. The QED BW process occurs in every polarization channel, albeit with significant variation in angle resolved DCS (see Supplementary Material I). In contrast, the ALP processes are restricted to very specific sets of polarizations. The pseudoscalar ALP requires that incoming linearly polarized photons have perpendicular polarization while the scalar only mediates the BW process for incoming photons which have parallel polarization. 

We have derived the cross section for the linear BW pair production mediated by an ALP which couples with photons and electrons. A central result is that the ALP-mediated channel should not be treated as an incoherent correction to the QED BW process. Instead, the ALP amplitude should be coherently combined with the QED amplitudes, as in Eq.~\eqref{eq:coherent}. Therefore both the magnitude of the ALP contribution and its phase relative to the QED continuum determine the precise cross section. For $s\simeq m_a^2$, the resulting interference replaces the ordinary Breit-Wigner resonance with a Fano-type profile, meaning that the cross section will exhibit a deficit of $e^+e^-$ pairs on one side of the resonance and an excess on the other, compared with the QED prediction.

For the benchmark cross section shown in Fig.~\ref{fig:xsection}, the difference between the QED prediction $\sigma_{\rm QED}$ and the coherently combined cross section $\sigma$ is
significant.  The presence or absence of an ALP $a(m_a,g_{a\gamma\gamma},g_{aee})$ in such a region of parameter space should be immediately apparent from a measurement of the linear BW process in the correct COM energy region. The benchmark should, however, be primarily interpreted as a demonstration of principle. For the minimal constant-coupling ALP EFT (considered above), the existence of ALPs with values of $g_{aee}$ much higher than about $10^{-5}$ are already under strong pressure from precision measurements of the electron anomalous magnetic moment $(g-2)$ \cite{Yan2019}. Thus, a role of the present calculation is, rather than to claim that this specific illustrative point in parameter space is viable, to show the manner in which the near threshold BW production will respond to a pseudoscalar ALP-mediated amplitude as the qualitative signature to be searched for experimentally.

The general ALP signature is more phenomenologically rich than a simple resonant bump. The ALP amplitude interfering with the QED continuum means that the observable distortion of QED BW depends on the COM energy spectrum of incoming photons, their polarizations and the angular distribution of produced $e^+e^-$ pairs. A scan over $\sqrt{s}$ in the vicinity of $m_a$ would therefore not only test the total rate, but also the line-shape, while polarization control would allow one to effectively switch on or off the ALP-mediated process. Therefore, polarization- and energy-resolved measurements could be especially valuable in providing the ability to distinguish between genuine ALP enhancements and normalization errors or background contributions. A scalar-mediated process has a very similar polarization averaged and spin summed cross section to the pseudoscalar-mediated process. However, it is suppressed near the pair production threshold by an additional $\beta^2$ factor (see Eq.~\eqref{eq:scalar_xsec} and Eq.~\eqref{eq:scalar_int} in Appendix D). Nonetheless, polarization resolved measurements would be exceptionally powerful in distinguishing between the contribution of a pseudoscalar or scalar ALP.

Throughout, we have made calculations within a low-energy EFT, treating $m_a$, $g_{a\gamma\gamma}$ and $g_{aee}$ as independent phenomenological parameters where, unlike the QCD axion, the ALP need not obey a fixed relationship between mass and coupling. Its interaction with Standard Model fields may be described at leading order by high-dimensional operators. For instance, the dimension-five $aF_{\mu\nu}\tilde{F}_{\mu\nu}$ coupling with the photon field is suppressed by an energy scale $\Lambda$, related to the physics which has been integrated out \cite{Bauer2021}. At low energy, such an EFT can permit predictive perturbative expansions. However, when the energies being probed become comparable to this energy scale, one needs to think carefully about whether a fixed coupling EFT is the correct language with which to describe the ALP interactions. Since the UV physics may start to be resolved at this energy, meaning that the effective interaction fails to fully account for the physics that has been ``integrated out", higher-dimensional operators might start to become more relevant. Additionally, the validity of perturbative treatments involving such a vertex should be called into question. Accordingly, the threshold BW process need not only be a probe of universal, energy-independent couplings. 

In principle, the microscopic dynamics generating the $a\gamma\gamma$ or $aee$ interaction may be better represented by effective couplings given by $s$-dependent form factors and the relevant quantity to measure is the magnitude of the ALP amplitude at particular values of COM energy. In this sense, cross section curves like those in Fig.~\ref{fig:xsection} can serve an additional purpose. Namely, detailing the response to an $s$-dependent coupling at different values of s. Therefore, scanning over COM energy in a linear BW experiment and comparing with QED effectively constrains the possible ALP form factors. A complete understanding of these constraints necessarily considers the consequences of generic complex valued form factors or propagator poles which could arise from the hypothetical microscopic physics, but this is left for future work. 

It is helpful to compare a near threshold BW search with existing ALP constraints and the approaches with which they have been derived. Helioscope \cite{CAST2017} and light-shining-through-wall \cite{Brotherton2026} experiments place extremely strong bounds on very light ALPs coupled to photons by probing coherent photon-ALP conversion at low energy. Experiments where accelerator beams are collided with a stationary target \cite{Riordan1987,Bergsma1985,Bjorken2009,Dobrich2016,Capozzi2023}, or nuclear reactor based experiments \cite{NEON2025}, can provide coverage in the keV - MeV mass regime. However, there are upper and lower limits to the coupling strengths which may be probed with these approaches. In particular, beam-dump experiments are sensitive to coupling lying in a ``sweet spot" that is high enough that ALP may be produced in sufficient quantity when the beam impinges upon the target and weak enough that the ALP doesn't just decay immediately within the ``dump".  Furthermore, astrophysical considerations such as bounding the stellar cooling which might occur assuming the existence of a \textit{sufficiently long-lived} ALP \cite{Hardy2017} place constraints on the parameter space; famously, observations from globular clusters in combination with the results of beam-dump experiments leave a small triangle of parameter space around $m_a\approx1$ MeV, whose exclusion depends upon choice of cosmological model \cite{Carenza2020}. Precision measurements of the electron $g-2$ \cite{Yan2019} and of positronium lifetimes \cite{Adkins2022} are also powerful, but they probe different regions of the amplitude to the linear BW process. Generally, $g-2$ measurements constrain loop-induced vertex corrections (in fact, through these effects, hypothetical ALP contributions have been shown to be capable of resolving the $g-2$ discrepancy in muons \cite{Marciano2016}), while positronium decay measurements probe the pole structure just below the $e^+e^-$ threshold. Finally, high-energy collider searches including $e^+e^-$ colliders \cite{Mimasu2015,Jaeckel2016}, ultra-peripheral heavy-ion collisions \cite{Knapen2017,STAR2021,ATLAS2017,ATLAS2021,CMS2025} and ``photon fusion" searches in proton colliders \cite{Atlas2020,Atlas2023} are sensitive at high masses, typically optimized for ALP searches in the hundreds-of-MeV to GeV regime, even up to thousands of GeV.
A near threshold BW measurement occupies a distinct and complementary position; it directly probes the timelike two-photon response of the vacuum at the opening of the real $e^+e^-$ channel. 

A common theme is that the exclusion plots generated by these experiments operate under the assumption of EFT validity and often restricted to ALPs which do not couple jointly to the photon and electron fields. Breaking from these assumptions will alter the conclusions which may be made regarding ALP parameter space exclusion. Firstly, the BW process can play a special role in ALP searches, since it necessarily constrains both $g_{aee}$ and $g_{a\gamma\gamma}$ simultaneously. Second, in order for the interference of ALP-mediated BW to cause a significant deviation from the QED BW prediction across a broad range in COM energy, one requires couplings at least of an order similar to those used in the examples shown in Fig.~\ref{fig:xsection}. At such high coupling, $\Lambda\sim\sqrt{s}\sim$ MeV, so the EFT picture may already be dubious. Such an ALP would offer the exciting possibility of probing its actual UV physics through BW measurements. If the ALP interacts with light through charged-particle ``triangle" loops, then their mass being nearby in energy would certainly create structure that is not captured by a constant-coupling EFT. For example, some non-trivial threshold behaviour in the form factors associated with the coupling would form if the ALP-photon coupling were induced by electrons themselves or, more speculatively, some millicharged fermion \cite{Davidson2000} with mass close to $m_e$. This possibility of energy dependent ALP couplings, and overall uncertainty in the expected cross section (without knowledge of the UV physics) illustrates that the BW pair production is best utilized as a probe of the $\gamma\gamma\to e^+e^-$ amplitude and its $s$-dependence. 

There are several theoretical refinements that should be addressed in future work. Very close to the $e^+e^-$ threshold, Coulomb/Sommerfeld effects \cite{Arbuzov2012} and the
nearby positronium spectrum \cite{Adkins2022} will modify the QED continuum line shape. Moreover, they may also modify the ALP amplitudes and these effect should be taken into account for a truly precise assessment of the near threshold BW process. 
In addition, as discussed in Appendix A we related the imaginary part of the ALP self-energy to the ALP decay width using the optical theorem. The real part of the self-energy has been neglected here. This real part shifts the location of the resonance and, close to threshold, may be sensitive to the UV completion. Again, a fully precise prediction of the resonant line shape may therefore require a microscopic model to be specified or a more general form-factor parameterization. 
These effects, aside from being theoretical complications, just add to the rich range of physics which may be probed by the BW process.

From the experimental modelling perspective, the relevant observable is the convolution of the fixed-$s$ cross section with the photon luminosity spectrum and polarization configuration on the source side and energy resolution and angular acceptance on the detector side. A narrow ALP resonance may be significantly smeared if the photon bandwidth is much wider than the Fano profile. Conversely, even a narrow structure might produce a substantial integrated deviation from QED if luminosity is sufficient in the vicinity of the resonance.
Null results in such measurements would therefore provide direct  exclusion of regions of the $(m_a,g_{a\gamma\gamma},g_{aee})$ parameter space depending on the luminosity spectrum and detector resolution. On the other hand, statistically significant deviations from the QED prediction, especially if combined with the expected line-shape, angular and polarization dependence, would constitute valuable evidence of a new ALP amplitude or, more broadly, non-standard MeV-scale dynamics in the two-photon channel.

In summary, precision measurements of linear BW pair production could provide a clean and unusually direct probe of MeV-scale BSM physics. For ALPs, they would test the interference of a possible $s$-channel ALP amplitude with the QED continuum. More generally, they probe an amplitude at a qualitative threshold: the point at which the real $e^+e^-$ continuum opens. This constitutes a different discovery mode from increasing collider energy or improving precision of constraints from static interactions and indeed probes a kinematic region of the amplitude not fully covered by helioscopes, light-shining-through-wall searches, beam dumps, reactor experiments, positronium spectroscopy, $g-2$ measurements, or at high-energy colliders.

Modern laser-plasma physics and the development of high-brightness photon sources has brought what was once  deemed ``hopeless" by Breit and Wheeler \cite{Breit1934}, laboratory exploration of the linear BW process with real photons, within experimental reach \cite{Kettle2021}. We anticipate near-term experimental realizations of the BW process,  either confirming QED or challenging it. 
Further improvements in photon flux, bandwidth control, polarimetry and pair detection could transform BW pair production into a precision tool for testing QED, as well as searching for ALPs and other BSM physics in the MeV regime.

\section{Acknowledgments}
All data and analyses are UK Ministry of Defence © Crown owned copyright 2026/AWE.
T.R. thanks B. King and G. Singh for interesting discussions about this work.

\FloatBarrier

\bibliography{ref}

@article{STAR2021,
  title = {Measurement of ${e}^{+}{e}^{\ensuremath{-}}$ Momentum and Angular Distributions from Linearly Polarized Photon Collisions},
  author = {{STAR Collaboration}},
  journal = {Phys. Rev. Lett.},
  volume = {127},
  issue = {5},
  pages = {052302},
  numpages = {9},
  year = {2021},
  month = {7},
  publisher = {American Physical Society},
  doi = {10.1103/PhysRevLett.127.052302},
  url = {https://link.aps.org/doi/10.1103/PhysRevLett.127.052302}
}

@article{ATLAS2017,
  title = {Evidence for light-by-light scattering in heavy-ion collisions with the $\mathrm{ATLAS}$ detector at the $\mathrm{LHC}$},
  author = {{ATLAS Collaboration}},
  journal = {Nature Physics},
  volume = {13},
  issue = {9},
  pages = {852--858},
  numpages = {7},
  year = {2017},
  month = {9},
  publisher = {Nature},
  doi = {10.1038/nphys4208},
  url = {https://doi.org/10.1038/nphys4208}
}

@article{ATLAS2021,
  title = {Measurement of light-by-light scattering and search for axion-like particles with $2.2 \textrm{nb}^{-1}$ of $\textrm{Pb}+\textrm{Pb}$ data with the $\mathrm{ATLAS}$ detector},
  author = {{ATLAS Collaboration}},
  journal = {Journal of High Energy Physics},
  volume = {2021},
  issue = {3},
  pages = {243},
  year = {2021},
  month = {3},
  publisher = {Springer},
  doi = {10.1007/JHEP03(2021)243},
  url = {https://doi.org/10.1007/JHEP03(2021)243}
}

@article{CMS2025,
  title = {Measurement of light-by-light scattering and the $\mathrm{Breit-Wheeler}$ process, and search for axion-like particles in ultraperipheral $\textrm{PbPb}$ collisions at $\sqrt{{s}_{\text{NN}}}$= 5.02 $\textrm{TeV}$},
  author = {{The CMS collaboration}},
  journal = {Journal of High Energy Physics},
  volume = {2025},
  issue = {8},
  pages = {6},
  year = {2025},
  month = {8},
  publisher = {Springer},
  doi = {10.1007/JHEP08(2025)006},
  url = {https://doi.org/10.1007/JHEP08(2025)006}
}

@article{Halverson2019,
  title = {Towards string theory expectations for photon couplings to axionlike particles},
  author = {Halverson, James and Long, Cody and Nelson, Brent and Salinas, Gustavo},
  journal = {Phys. Rev. D},
  volume = {100},
  issue = {10},
  pages = {106010},
  numpages = {13},
  year = {2019},
  month = {11},
  publisher = {American Physical Society},
  doi = {10.1103/PhysRevD.100.106010},
  url = {https://link.aps.org/doi/10.1103/PhysRevD.100.106010}
}

@article{Brivio2021,
  title = {Unitarity constraints on $\mathrm{ALP}$ interactions},
  author = {Brivio, I. and \'Eboli, O. J. P. and Gonz\'alez-Garc\'{\i}a, M. C.},
  journal = {Phys. Rev. D},
  volume = {104},
  issue = {3},
  pages = {035027},
  numpages = {11},
  year = {2021},
  month = {8},
  publisher = {American Physical Society},
  doi = {10.1103/PhysRevD.104.035027},
  url = {https://link.aps.org/doi/10.1103/PhysRevD.104.035027}
}

@article{Bresciani2025A,
  title = {Amplitudes and partial wave unitarity bounds},
  author = {Bresciani, Luigi C. and Levati, Gabriele and Paradisi, Paride},
  journal = {Phys. Rev. D},
  volume = {113},
  issue = {7},
  pages = {L071702},
  numpages = {7},
  year = {2026},
  month = {4},
  publisher = {American Physical Society},
  doi = {10.1103/tp6m-mcyn},
  url = {https://link.aps.org/doi/10.1103/tp6m-mcyn}
}

@ARTICLE{Bresciani2025B,
       author = {{Bresciani}, Luigi C. and {Levati}, Gabriele and {Paradisi}, Paride},
        title = "{Positivity and partial wave unitarity bounds on ALP theories via amplitude methods}",
      journal = {arXiv e-prints},
         year = 2025,
        month = 10,
archivePrefix = {arXiv},
       eprint = {2510.13953},
 primaryClass = {hep-ph},
       adsurl = {https://ui.adsabs.harvard.edu/abs/2025arXiv251013953B},
}

@article{Esnault2021,
doi = {10.1088/1361-6587/ac2e3e},
url = {https://doi.org/10.1088/1361-6587/ac2e3e},
year = {2021},
month = {11},
publisher = {IOP Publishing},
volume = {63},
number = {12},
pages = {125015},
author = {Esnault, L and d’Humières, E and Arefiev, A and Ribeyre, X},
title = {Electron-positron pair production in the collision of real photon beams with wide energy distributions},
journal = {Plasma Physics and Controlled Fusion}
}

@article{Kettle2021,
doi = {10.1088/1367-2630/ac3048},
url = {https://doi.org/10.1088/1367-2630/ac3048},
year = {2021},
month = {11},
publisher = {IOP Publishing},
volume = {23},
number = {11},
pages = {115006},
author = {Kettle, B and Hollatz, D and Gerstmayr, E and others},
title = {A laser–plasma platform for photon–photon physics: the two photon $\mathrm{Breit-Wheeler}$ process},
journal = {New Journal of Physics}
}

@article{Watt2025,
title = {Bounding elastic photon-photon scattering at $\sqrt{s}\approx 1 \mathrm{MeV}$ using a laser-plasma platform},
journal = {Physics Letters B},
volume = {861},
pages = {139247},
year = {2025},
issn = {0370-2693},
doi = {https://doi.org/10.1016/j.physletb.2025.139247},
url = {https://www.sciencedirect.com/science/article/pii/S0370269325000073},
author = {R. Watt and B. Kettle and E. Gerstmayr and others}
}

@article{Galanti2022,
doi = {10.3390/universe8050253},
url = {https://www.mdpi.com/2218-1997/8/5/253},
year = {2022},
month = {5},
publisher = {MDPI},
volume = {8},
number = {5},
pages = {253},
author = {Galanti, Giorgio and Roncadelli, Marco},
title = {Axion-like Particles Implications for High-Energy Astrophysics},
journal = {Universe}
}

@article{Breit1934,
  title = {Collision of Two Light Quanta},
  author = {Breit, G. and Wheeler, John A.},
  journal = {Phys. Rev.},
  volume = {46},
  issue = {12},
  pages = {1087--1091},
  numpages = {0},
  year = {1934},
  month = {12},
  publisher = {American Physical Society},
  doi = {10.1103/PhysRev.46.1087},
  url = {https://link.aps.org/doi/10.1103/PhysRev.46.1087}
}

@article{Eichhorn2012,
  title = {Renormalization flow of axion electrodynamics},
  author = {Eichhorn, Astrid and Gies, Holger and Roscher, Dietrich},
  journal = {Phys. Rev. D},
  volume = {86},
  issue = {12},
  pages = {125014},
  numpages = {12},
  year = {2012},
  month = {12},
  publisher = {American Physical Society},
  doi = {10.1103/PhysRevD.86.125014},
  url = {https://link.aps.org/doi/10.1103/PhysRevD.86.125014}
}

@article{Dupuis2021,
title = {The nonperturbative functional renormalization group and its applications},
journal = {Physics Reports},
volume = {910},
pages = {1-114},
year = {2021},
issn = {0370-1573},
doi = {https://doi.org/10.1016/j.physrep.2021.01.001},
url = {https://www.sciencedirect.com/science/article/pii/S0370157321000156},
author = {N. Dupuis and L. Canet and A. Eichhorn and W. Metzner and J.M. Pawlowski and M. Tissier and N. Wschebor}
}

@article{deBrito2022,
  title = {Are there $\mathrm{ALP}$s in the asymptotically safe landscape?},
  author = {de Brito, Gustavo P. and Eichhorn, Astrid and Lino dos Santos, Rafael R.},
  journal = {Journal of High Energy Physics},
  volume = {2022},
  issue = {6},
  pages = {13},
  numpages = {33},
  year = {2022},
  month = {6},
  publisher = {Springer},
  doi = {10.1007/JHEP06(2022)013},
  url = {https://doi.org/10.1007/JHEP06(2022)013}
}

@article{Voisin2017,
    author = {Voisin, Guillaume and Mottez, Fabrice and Bonazzola, Silvano},
    title = {Electron–positron pair production by gamma-rays in an anisotropic flux of soft photons, and application to pulsar polar caps},
    journal = {Monthly Notices of the Royal Astronomical Society},
    volume = {474},
    number = {2},
    pages = {1436-1452},
    year = {2017},
    month = {10},
    issn = {0035-8711},
    doi = {10.1093/mnras/stx2658},
    url = {https://doi.org/10.1093/mnras/stx2658},
}

@article{Cabral2023,
    author = {Cabral, D. S. and Santos, A. F. and Bufalo, R.},
    title = {Thermal pair production from photon-photon collision: $\mathrm{Breit-Wheeler}$ process at finite temperature},
    journal = {The European Physical Journal C},
    volume = {83},
    number = {12},
    pages = {1113-1121},
    year = {2023},
    month = {12},
    issn = {1434-6052},
    doi = {10.1140/epjc/s10052-023-12281-5},
    url = {https://doi.org/10.1140/epjc/s10052-023-12281-5}
}

@article{Song2025,
doi = {10.1088/1367-2630/ade61c},
url = {https://doi.org/10.1088/1367-2630/ade61c},
year = {2025},
month = {jul},
publisher = {IOP Publishing},
volume = {27},
number = {7},
pages = {074301},
author = {Song, Huai-Hang and Sheng, Zheng-Ming},
title = {Photon-polarization-resolved linear $\mathrm{Breit-Wheeler}$ pair production in a laser-plasma system},
journal = {New Journal of Physics}
}

@Inbook{Peccei2008,
author="Peccei, Roberto D.",
editor="Kuster, Markus
and Raffelt, Georg
and Beltr{\'a}n, Berta",
title="The Strong $\mathrm{CP}$ Problem and Axions",
bookTitle="Axions: Theory, Cosmology, and Experimental Searches",
year="2008",
publisher="Springer Berlin Heidelberg",
address="Berlin, Heidelberg",
pages="3--17",
isbn="978-3-540-73518-2",
doi="10.1007/978-3-540-73518-2_1",
url="https://doi.org/10.1007/978-3-540-73518-2_1"
}

@article{Kim2010,
  title = {Axions and the strong $\mathrm{CP}$ problem},
  author = {Kim, Jihn E. and Carosi, Gianpaolo},
  journal = {Rev. Mod. Phys.},
  volume = {82},
  issue = {1},
  pages = {557--601},
  numpages = {0},
  year = {2010},
  month = {3},
  publisher = {American Physical Society},
  doi = {10.1103/RevModPhys.82.557},
  url = {https://link.aps.org/doi/10.1103/RevModPhys.82.557}
}

@article{Peccei1977a,
  title = {$\mathrm{CP}$ Conservation in the Presence of Pseudoparticles},
  author = {Peccei, R. D. and Quinn, Helen R.},
  journal = {Phys. Rev. Lett.},
  volume = {38},
  issue = {25},
  pages = {1440--1443},
  numpages = {0},
  year = {1977},
  month = {6},
  publisher = {American Physical Society},
  doi = {10.1103/PhysRevLett.38.1440},
  url = {https://link.aps.org/doi/10.1103/PhysRevLett.38.1440}
}

@article{Peccei1977b,
  title = {Constraints imposed by $\mathrm{CP}$ conservation in the presence of pseudoparticles},
  author = {Peccei, R. D. and Quinn, Helen R.},
  journal = {Phys. Rev. D},
  volume = {16},
  issue = {6},
  pages = {1791--1797},
  numpages = {0},
  year = {1977},
  month = {9},
  publisher = {American Physical Society},
  doi = {10.1103/PhysRevD.16.1791},
  url = {https://link.aps.org/doi/10.1103/PhysRevD.16.1791}
}

@article{Preskill1983,
title = {Cosmology of the invisible axion},
journal = {Physics Letters B},
volume = {120},
number = {1},
pages = {127-132},
year = {1983},
issn = {0370-2693},
doi = {https://doi.org/10.1016/0370-2693(83)90637-8},
url = {https://www.sciencedirect.com/science/article/pii/0370269383906378},
author = {John Preskill and Mark B. Wise and Frank Wilczek}
}

@book{Jauch1976,
    author = "Jauch, J. M. and Rohrlich, F.",
    title = "{The theory of photons and electrons. The relativistic quantum field theory of charged particles with spin one-half}",
    edition = "2nd ed.",
    doi = "10.1007/978-3-642-80951-4",
    isbn = "978-3-642-80953-8, 978-3-642-80951-4",
    publisher = "Springer",
    address = "Berlin",
    series = "Texts and Monographs in Physics",
    year = "1976"
}

@misc{AxionLimits,
  author       = {Ciaran O'Hare},
  title        = {cajohare/AxionLimits: AxionLimits},
  month        = jul,
  year         = 2020,
  publisher    = {Zenodo},
  version      = {v1.0},
  doi          = {10.5281/zenodo.3932430},
  howpublished = {\url{https://cajohare.github.io/AxionLimits/}}
}

@article{Yan2019,
title = {Constraining exotic spin dependent interactions of muons and electrons},
journal = {The European Physical Journal C},
volume = {79},
number = {11},
pages = {971},
year = {2019},
issn = {1434-6052},
doi = {10.1140/epjc/s10052-019-7442-8},
url = {https://doi.org/10.1140/epjc/s10052-019-7442-8},
author = {Yan, H. and Sun, G. A. and Peng, S. M. and Guo, H. and Liu, B. Q. and Peng, M. and Zheng, H.}
}

@article{Yamada2021,
title = {A natural and simple UV completion of the QCD axion model},
journal = {Physics Letters B},
volume = {816},
pages = {136267},
year = {2021},
issn = {0370-2693},
doi = {https://doi.org/10.1016/j.physletb.2021.136267},
url = {https://www.sciencedirect.com/science/article/pii/S0370269321002070},
author = {Masaki Yamada and Tsutomu T. Yanagida}
}

@article{Dine1981,
title = {A simple solution to the strong $\mathrm{CP}$ problem with a harmless axion},
journal = {Physics Letters B},
volume = {104},
number = {3},
pages = {199-202},
year = {1981},
issn = {0370-2693},
doi = {https://doi.org/10.1016/0370-2693(81)90590-6},
url = {https://www.sciencedirect.com/science/article/pii/0370269381905906},
author = {Michael Dine and Willy Fischler and Mark Srednicki}
}

@article{Shifman1980,
title = {Can confinement ensure natural $\mathrm{CP}$ invariance of strong interactions?},
journal = {Nuclear Physics B},
volume = {166},
number = {3},
pages = {493-506},
year = {1980},
issn = {0550-3213},
doi = {https://doi.org/10.1016/0550-3213(80)90209-6},
url = {https://www.sciencedirect.com/science/article/pii/0550321380902096},
author = {M.A. Shifman and A.I. Vainshtein and V.I. Zakharov}
}

@article{Redondo2011,
author = {Javier Redondo and Andreas Ringwald},
title = {Light shining through walls},
journal = {Contemporary Physics},
volume = {52},
number = {3},
pages = {211--236},
year = {2011},
publisher = {Taylor \& Francis},
doi = {10.1080/00107514.2011.563516},
URL = {https://doi.org/10.1080/00107514.2011.563516},
}

@article{Sikivie1983,
  title = {Experimental Tests of the "Invisible" Axion},
  author = {Sikivie, P.},
  journal = {Phys. Rev. Lett.},
  volume = {51},
  issue = {16},
  pages = {1415--1417},
  numpages = {0},
  year = {1983},
  month = {10},
  publisher = {American Physical Society},
  doi = {10.1103/PhysRevLett.51.1415},
  url = {https://link.aps.org/doi/10.1103/PhysRevLett.51.1415}
}

@article{Pike2014,
  title = {A photon–photon collider in a vacuum hohlraum},
  author = {Pike, O. J. and Mackenroth, F. and Hill, E. G. and Rose, S. J.},
  journal = {Nature Photonics},
  volume = {8},
  issue = {6},
  pages = {434-436},
  numpages = {3},
  year = {2014},
  month = {6},
  publisher = {Springer Nature},
  doi = {10.1038/nphoton.2014.95},
  url = {https://doi.org/10.1038/nphoton.2014.95}
}

@article{Bethe1947,
  title = {The Electromagnetic Shift of Energy Levels},
  author = {Bethe, H. A.},
  journal = {Phys. Rev.},
  volume = {72},
  issue = {4},
  pages = {339--341},
  numpages = {0},
  year = {1947},
  month = {8},
  publisher = {American Physical Society},
  doi = {10.1103/PhysRev.72.339},
  url = {https://link.aps.org/doi/10.1103/PhysRev.72.339}
}

@article{Dirac1927,
    author = {Dirac, Paul Adrien Maurice},
    title = {The quantum theory of the emission and absorption of radiation},
    journal = {Proceedings of the Royal Society of London. Series A, Containing Papers of a Mathematical and Physical Character},
    volume = {114},
    number = {767},
    pages = {243-265},
    year = {1927},
    month = {03},
    issn = {0950-1207},
    doi = {10.1098/rspa.1927.0039},
    url = {https://doi.org/10.1098/rspa.1927.0039},
}

@article{Hanneke2008,
  title = {New Measurement of the Electron Magnetic Moment and the Fine Structure Constant},
  author = {Hanneke, D. and Fogwell, S. and Gabrielse, G.},
  journal = {Phys. Rev. Lett.},
  volume = {100},
  issue = {12},
  pages = {120801},
  numpages = {4},
  year = {2008},
  month = {3},
  publisher = {American Physical Society},
  doi = {10.1103/PhysRevLett.100.120801},
  url = {https://link.aps.org/doi/10.1103/PhysRevLett.100.120801}
}

@article{Aoyama2012,
  title = {Tenth-Order $\mathrm{QED}$ Contribution to the Electron $g\mathbf{\ensuremath{-}}2$ and an Improved Value of the Fine Structure Constant},
  author = {Aoyama, Tatsumi and Hayakawa, Masashi and Kinoshita, Toichiro and Nio, Makiko},
  journal = {Phys. Rev. Lett.},
  volume = {109},
  issue = {11},
  pages = {111807},
  numpages = {4},
  year = {2012},
  month = {Sep},
  publisher = {American Physical Society},
  doi = {10.1103/PhysRevLett.109.111807},
  url = {https://link.aps.org/doi/10.1103/PhysRevLett.109.111807}
}

@article{Kusch1948,
  title = {The Magnetic Moment of the Electron},
  author = {Kusch, P. and Foley, H. M.},
  journal = {Phys. Rev.},
  volume = {74},
  issue = {3},
  pages = {250--263},
  numpages = {0},
  year = {1948},
  month = {8},
  publisher = {American Physical Society},
  doi = {10.1103/PhysRev.74.250},
  url = {https://link.aps.org/doi/10.1103/PhysRev.74.250}
}

@article{Lamb1947,
  title = {Fine Structure of the Hydrogen Atom by a Microwave Method},
  author = {Lamb, Willis E. and Retherford, Robert C.},
  journal = {Phys. Rev.},
  volume = {72},
  issue = {3},
  pages = {241--243},
  numpages = {0},
  year = {1947},
  month = {8},
  publisher = {American Physical Society},
  doi = {10.1103/PhysRev.72.241},
  url = {https://link.aps.org/doi/10.1103/PhysRev.72.241}
}

@article{Dyson1949,
  title = {The Radiation Theories of Tomonaga, Schwinger, and Feynman},
  author = {Dyson, F. J.},
  journal = {Phys. Rev.},
  volume = {75},
  issue = {3},
  pages = {486--502},
  numpages = {0},
  year = {1949},
  month = {2},
  publisher = {American Physical Society},
  doi = {10.1103/PhysRev.75.486},
  url = {https://link.aps.org/doi/10.1103/PhysRev.75.486}
}

@article{Feynman1949,
  title = {Space-Time Approach to Quantum Electrodynamics},
  author = {Feynman, R. P.},
  journal = {Phys. Rev.},
  volume = {76},
  issue = {6},
  pages = {769--789},
  numpages = {0},
  year = {1949},
  month = {9},
  publisher = {American Physical Society},
  doi = {10.1103/PhysRev.76.769},
  url = {https://link.aps.org/doi/10.1103/PhysRev.76.769}
}

@article{Tomonga1946,
    author = {Tomonaga, S.},
    title = {On a Relativistically Invariant Formulation of the Quantum Theory of Wave Fields*},
    journal = {Progress of Theoretical Physics},
    volume = {1},
    number = {2},
    pages = {27-42},
    year = {1946},
    month = {08},
    issn = {0033-068X},
    doi = {10.1143/PTP.1.27},
    url = {https://doi.org/10.1143/PTP.1.27},
}

@article{Schwinger1948,
  title = {Quantum Electrodynamics. I. A Covariant Formulation},
  author = {Schwinger, Julian},
  journal = {Phys. Rev.},
  volume = {74},
  issue = {10},
  pages = {1439--1461},
  numpages = {0},
  year = {1948},
  month = {11},
  publisher = {American Physical Society},
  doi = {10.1103/PhysRev.74.1439},
  url = {https://link.aps.org/doi/10.1103/PhysRev.74.1439}
}

@article{Dent2020,
  title = {New Directions for Axion Searches via Scattering at Reactor Neutrino Experiments},
  author = {Dent, James B. and Dutta, Bhaskar and Kim, Doojin and Liao, Shu and Mahapatra, Rupak and Sinha, Kuver and Thompson, Adrian},
  journal = {Phys. Rev. Lett.},
  volume = {124},
  issue = {21},
  pages = {211804},
  numpages = {6},
  year = {2020},
  month = {5},
  publisher = {American Physical Society},
  doi = {10.1103/PhysRevLett.124.211804},
  url = {https://link.aps.org/doi/10.1103/PhysRevLett.124.211804}
}

@article{Wong2021,
  title = {Dark photon production via $\gamma \gamma \rightarrow \gamma A'$},
  author = {Wong, Xiaorui and Huang, Yongsheng},
  journal = {The European Physical Journal C},
  volume = {81},
  issue = {5},
  pages = {442},
  numpages = {6},
  year = {2021},
  month = {5},
  publisher = {Springer},
  doi = {10.1140/epjc/s10052-021-09228-z},
  url = {https://doi.org/10.1140/epjc/s10052-021-09228-z}
}

@article{Davidson2000,
doi = {10.1088/1126-6708/2000/05/003},
url = {https://doi.org/10.1088/1126-6708/2000/05/003},
year = {2000},
month = {5},
publisher = {},
volume = {2000},
number = {05},
pages = {003},
author = {Sacha Davidson and Steen Hannestad and Georg Raffelt},
title = {Updated bounds on milli-charged particles},
journal = {Journal of High Energy Physics}
}

@article{Nikishov1962,
    author = "Nikishov, A. I.",
    title = "{Absorption of High-Energy Photons in the Universe}",
    journal = "Sov. Phys. JETP",
    volume = "14",
    pages = "393--394",
    year = "1962"
}

@article{Gould1967,
  title = {Opacity of the Universe to High-Energy Photons},
  author = {Gould, Robert J. and Schr\'eder, G\'erard P.},
  journal = {Phys. Rev.},
  volume = {155},
  issue = {5},
  pages = {1408--1411},
  numpages = {0},
  year = {1967},
  month = {3},
  publisher = {American Physical Society},
  doi = {10.1103/PhysRev.155.1408},
  url = {https://link.aps.org/doi/10.1103/PhysRev.155.1408}
}

@article{Atlas2020,
  title = {Observation and Measurement of Forward Proton Scattering in Association with Lepton Pairs Produced via the Photon Fusion Mechanism at $\mathrm{ATLAS}$},
  author = {{ATLAS Collaboration}},
  journal = {Phys. Rev. Lett.},
  volume = {125},
  issue = {26},
  pages = {261801},
  numpages = {21},
  year = {2020},
  month = {12},
  publisher = {American Physical Society},
  doi = {10.1103/PhysRevLett.125.261801},
  url = {https://link.aps.org/doi/10.1103/PhysRevLett.125.261801}
}

@article{Atlas2023,
  title = {Search for an axion-like particle with forward proton scattering in association with photon pairs at $\mathrm{ATLAS}$},
  author = {{ATLAS Collaboration}},
  journal = {Journal of High Energy Physics},
  volume = {2023},
  issue = {7},
  pages = {234},
  numpages = {21},
  year = {2023},
  month = {12},
  publisher = {Springer},
  doi = {10.1007/JHEP07(2023)234},
  url = {https://doi.org/10.1007/JHEP07(2023)234}
}

@article{Marciano2016,
  title = {Contributions of axionlike particles to lepton dipole moments},
  author = {Marciano, W. J. and Masiero, A. and Paradisi, P. and Passera, M.},
  journal = {Phys. Rev. D},
  volume = {94},
  issue = {11},
  pages = {115033},
  numpages = {7},
  year = {2016},
  month = {12},
  publisher = {American Physical Society},
  doi = {10.1103/PhysRevD.94.115033},
  url = {https://link.aps.org/doi/10.1103/PhysRevD.94.115033}
}

@article{CAST2017,
  title = {New $\mathrm{CAST}$ limit on the axion–photon interaction},
  author = {{CAST Collaboration}},
  journal = {Nature Physics},
  volume = {13},
  issue = {6},
  pages = {584590},
  numpages = {7},
  year = {2017},
  month = {06},
  publisher = {Springer},
  doi = {10.1038/nphys4109},
  url = {https://doi.org/10.1038/nphys4109}
}

@misc{Brotherton2026,
      title={Any Light Particle Searches with $\mathrm{ALP}$ II: first science results}, 
      author={Daniel C. Brotherton and Zachary R. Bush and Sandy Croatto and others},
      year={2026},
      eprint={2512.14110},
      archivePrefix={arXiv},
      primaryClass={hep-ex},
      url={https://arxiv.org/abs/2512.14110}, 
}

@article{Bjorken2009,
  title = {New fixed-target experiments to search for dark gauge forces},
  author = {Bjorken, James D. and Essig, Rouven and Schuster, Philip and Toro, Natalia},
  journal = {Phys. Rev. D},
  volume = {80},
  issue = {7},
  pages = {075018},
  numpages = {20},
  year = {2009},
  month = {Oct},
  publisher = {American Physical Society},
  doi = {10.1103/PhysRevD.80.075018},
  url = {https://link.aps.org/doi/10.1103/PhysRevD.80.075018}
}

@article{Dobrich2016,
  title = {$\mathrm{ALP}$traum: $\mathrm{ALP}$ production in proton beam dump experiments},
  author = {Döbrich, Babette and Jaeckel, Joerg and Kahlhoefer, Felix and others},
  journal = {Journal of High Energy Physics},
  volume = {2016},
  issue = {2},
  pages = {18},
  numpages = {21},
  year = {2016},
  month = {02},
  publisher = {Springer},
  doi = {10.1007/JHEP02(2016)018},
  url = {https://doi.org/10.1007/JHEP02(2016)018}
}

@article{Riordan1987,
  title = {Search for short-lived axions in an electron-beam-dump experiment},
  author = {Riordan, E. M. and Krasny, M. W. and Lang, K. and others},
  journal = {Phys. Rev. Lett.},
  volume = {59},
  issue = {7},
  pages = {755--758},
  numpages = {0},
  year = {1987},
  month = {08},
  publisher = {American Physical Society},
  doi = {10.1103/PhysRevLett.59.755},
  url = {https://link.aps.org/doi/10.1103/PhysRevLett.59.755}
}

@article{Bergsma1985,
title = {Search for axion-like particle production in 400 $\mathrm{GeV}$ proton-copper interactions},
journal = {Physics Letters B},
volume = {157},
number = {5},
pages = {458-462},
year = {1985},
issn = {0370-2693},
doi = {https://doi.org/10.1016/0370-2693(85)90400-9},
url = {https://www.sciencedirect.com/science/article/pii/0370269385904009},
author = {F. Bergsma and J. Dorenbosch and J.V. Allaby and others},
}

@article{NEON2025,
  title = {New Constraints on Axionlike Particles with the NEON Detector at a Nuclear Reactor},
  author = {{NEON Collaboration}},
  journal = {Phys. Rev. Lett.},
  volume = {134},
  issue = {20},
  pages = {201002},
  numpages = {8},
  year = {2025},
  month = {May},
  publisher = {American Physical Society},
  doi = {10.1103/PhysRevLett.134.201002},
  url = {https://link.aps.org/doi/10.1103/PhysRevLett.134.201002}
}

@article{Capozzi2023,
  title = {New constraints on $\mathrm{ALP}$ couplings to electrons and photons from ArgoNeuT and the MiniBooNE beam dump},
  author = {Capozzi, Francesco and Dutta, Bhaskar and Gurung, Gajendra and others},
  journal = {Phys. Rev. D},
  volume = {108},
  issue = {7},
  pages = {075019},
  numpages = {9},
  year = {2023},
  month = {10},
  publisher = {American Physical Society},
  doi = {10.1103/PhysRevD.108.075019},
  url = {https://link.aps.org/doi/10.1103/PhysRevD.108.075019}
}

@article{Adkins2022,
title = {Precision spectroscopy of positronium: Testing bound-state $\mathrm{QED}$ theory and the search for physics beyond the Standard Model},
journal = {Physics Reports},
volume = {975},
pages = {1-61},
year = {2022},
issn = {0370-1573},
doi = {https://doi.org/10.1016/j.physrep.2022.05.002},
url = {https://www.sciencedirect.com/science/article/pii/S0370157322001508},
author = {G.S. Adkins and D.B. Cassidy and J. Pérez-Ríos},
}

@article{Jaeckel2016,
title = {Probing $\mathrm{MeV}$ to 90 $\mathrm{GeV}$ axion-like particles with $\mathrm{LEP}$ and $\mathrm{LHC}$},
journal = {Physics Letters B},
volume = {753},
pages = {482-487},
year = {2016},
issn = {0370-2693},
doi = {https://doi.org/10.1016/j.physletb.2015.12.037},
url = {https://www.sciencedirect.com/science/article/pii/S0370269315009855},
author = {Joerg Jaeckel and Michael Spannowsky},
}

@article{Mimasu2015,
  title = {$\mathrm{ALP}$s at colliders},
  author = {Mimasu, Ken and Sanz, Verónica},
  journal = {Journal of High Energy Physics},
  volume = {2015},
  issue = {6},
  pages = {173},
  numpages = {9},
  year = {2015},
  month = {06},
  publisher = {Springer},
  doi = {10.1007/JHEP06(2015)173},
  url = {https://doi.org/10.1007/JHEP06(2015)173}
}

@article{Knapen2017,
  title = {Searching for Axionlike Particles with Ultraperipheral Heavy-Ion Collisions},
  author = {Knapen, Simon and Lin, Tongyan and Lou, Hou Keong and Melia, Tom},
  journal = {Phys. Rev. Lett.},
  volume = {118},
  issue = {17},
  pages = {171801},
  numpages = {6},
  year = {2017},
  month = {4},
  publisher = {American Physical Society},
  doi = {10.1103/PhysRevLett.118.171801},
  url = {https://link.aps.org/doi/10.1103/PhysRevLett.118.171801}
}

@article{Arbuzov2012,
  title = {On relativization of the Sommerfeld-Gamow-Sakharov factor},
  author = {Arbuzov, Andrej B. and Kopylova, Tatiana V.},
  journal = {Journal of High Energy Physics},
  volume = {2012},
  issue = {4},
  pages = {9},
  numpages = {9},
  year = {2012},
  month = {04},
  publisher = {Springer},
  doi = {10.1007/JHEP04(2012)009},
  url = {https://doi.org/10.1007/JHEP04(2012)009}
}

@article{Bauer2021,
  title = {The low-energy effective theory of axions and $\mathrm{ALP}$s},
  author = {Bauer, Martin and Neubert, Matthias and Renner, Sophie and others},
  journal = {Journal of High Energy Physics},
  volume = {2021},
  issue = {4},
  pages = {63},
  numpages = {9},
  year = {2021},
  month = {04},
  publisher = {Springer},
  doi = {10.1007/JHEP04(2021)063},
  url = {https://doi.org/10.1007/JHEP04(2021)063}
}

@article{Hardy2017,
  title = {Stellar cooling bounds on new light particles: plasma mixing effects},
  author = {Hardy, Edward and Lasenby, Robert},
  journal = {Journal of High Energy Physics},
  volume = {2017},
  issue = {2},
  pages = {33},
  numpages = {9},
  year = {2017},
  month = {02},
  publisher = {Springer},
  doi = {10.1007/JHEP02(2017)033},
  url = {https://doi.org/10.1007/JHEP02(2017)033}
}

@article{Carenza2020,
title = {Constraints on the coupling with photons of heavy axion-like-particles from Globular Clusters},
journal = {Physics Letters B},
volume = {809},
pages = {135709},
year = {2020},
issn = {0370-2693},
doi = {https://doi.org/10.1016/j.physletb.2020.135709},
url = {https://www.sciencedirect.com/science/article/pii/S0370269320305128},
author = {Pierluca Carenza and Oscar Straniero and Babette Döbrich and Maurizio Giannotti and Giuseppe Lucente and Alessandro Mirizzi},
}

@article{Fano1961,
  title = {Effects of Configuration Interaction on Intensities and Phase Shifts},
  author = {Fano, U.},
  journal = {Phys. Rev.},
  volume = {124},
  issue = {6},
  pages = {1866--1878},
  numpages = {0},
  year = {1961},
  month = {12},
  publisher = {American Physical Society},
  doi = {10.1103/PhysRev.124.1866},
  url = {https://link.aps.org/doi/10.1103/PhysRev.124.1866}
}

@article{Breit1936,
  title = {Capture of Slow Neutrons},
  author = {Breit, G. and Wigner, E.},
  journal = {Phys. Rev.},
  volume = {49},
  issue = {7},
  pages = {519--531},
  numpages = {0},
  year = {1936},
  month = {4},
  publisher = {American Physical Society},
  doi = {10.1103/PhysRev.49.519},
  url = {https://link.aps.org/doi/10.1103/PhysRev.49.519}
}

@article{Davila2014,
author = {D\'{a}vila, Jos\'{e} Manuel and Schubert, Christian and Trejo, Mar\'{\i}a Anabel},
title = {Photonic processes in Born–Infeld theory},
journal = {International Journal of Modern Physics A},
volume = {29},
number = {30},
pages = {1450174},
year = {2014},
doi = {10.1142/S0217751X14501747},
URL = {https://doi.org/10.1142/S0217751X14501747}
}

\onecolumngrid

\setcounter{equation}{0}
\renewcommand{\theequation}{A\arabic{equation}}
\section*{Appendix}

\subsection{A. Pseudoscalar ALP-mediated Breit-Wheeler unpolarized cross section}
The ALP-QED Lagrangian [Eq.~\eqref{eq:Lagrange} in main text] has the additional Feynman rules for the:
\begin{itemize}
    \item ALP-photon vertex: $-ig_{a\gamma\gamma}k^{(1)}_\mu k^{(2)}_\lambda \epsilon^{\mu\nu\lambda\rho}$
    \item ALP-fermion vertex: $g_{aee}\gamma^5$
    \item ALP propagator: $\frac{i}{q^2-m_a^2-\Pi(s)}$ 
\end{itemize}
where $\Pi(s)$ is the ALP self-energy. These lead to the amplitude
\begin{equation}
\begin{split}
    \mathcal{M}_a^{(ijkl)} = g_{a\gamma\gamma}g_{aee}\epsilon^{(1)i}_\nu \epsilon^{(2)j}_\rho k^{(1)}_\mu k^{(2)}_\lambda \epsilon^{\mu\nu\lambda\rho} \frac{1}{(k^{(1)}+k^{(2)})^2 - m_a^2 +i\sqrt{s}\Gamma(s)} \bar{u}^{k}(p_1)\gamma^5 v^{l}(p_2) 
\end{split}
\end{equation}
for the Feynman diagram in Figure~\ref{fig:xsection}, that describes the pseudoscalar-ALP-mediated BW process.

In the absence of any information about the spins of outgoing electron-positron pairs and polarization of incoming photons, one must sum over the possible final spin states and average over the incoming polarizations.
To do this, we can use the fact that in the COM frame,
\begin{equation}\label{eq:spinSum}
\sum_{k,l}\bar{u}^{k}\gamma^5 v ^{l}\left[\bar{u}^{k}\gamma^5 v ^{l}\right]^\star = 2s,
\end{equation}
and
\begin{equation}\label{eq:polSum}
\begin{split}
    \frac{1}{4}\sum_{i,j}  \epsilon^{(1)i}_\nu \epsilon^{(2)j}_\rho k^{(1)}_\mu k^{(2)}_\lambda \epsilon^{\mu\nu\lambda\rho} \left[\epsilon^{(1)i}_\nu \epsilon^{(2)j}_\rho k^{(1)}_\mu k^{(2)}_\lambda \epsilon^{\mu\nu\lambda\rho}\right]^\star = \frac{s^2}{8}.
\end{split}
\end{equation}

So, the polarization averaged, spin summed amplitude squared will be 
\begin{equation}
    \langle\left|\mathcal{M}\right|^2\rangle = \frac{(g_{a\gamma\gamma}g_{aee})^2 s^3}{4|s - m_a^2 +i\sqrt{s}\Gamma(s)|^2}
\end{equation}
which, since $d\sigma/d\Omega = \left|M\right|^2 \sqrt{s-4m_e^2}/(64\pi^2 s \sqrt{s})$,
gives
\begin{equation}\label{eq:axion_crossSection}
    \sigma_a = \frac{(g_{a\gamma\gamma}g_{aee})^2}{64\pi}\frac{ s\sqrt{s}\sqrt{s-4m_e^2}}{|s - m_a^2 +i\sqrt{s}\Gamma(s)|^2}.
\end{equation}

At high energy, the cross section for the ALP-mediated BW process given by Eq.~\eqref{eq:axion_crossSection} is not expected to remain physical; the derivative coupling in Eq.~\eqref{eq:Lagrange}, between the ALP and photon field is typical of an effective coupling and it is unreasonable to expect this picture to be uniformly applicable across all energy scales. Understanding exactly how the physics changes at high energy, requires knowledge of the precise nature of the physics that was integrated out at lower energy. Some examples of possible UV completions come from QCD \cite{Peccei1977a,Peccei1977b,Shifman1980,Dine1981,Yamada2021} and string theory \cite{Halverson2019}, but the general parameter space is limitless. We have focused on the low energy region with the assumption that we are safely below the energy threshold, $\Lambda_\mathrm{UV}$, where a more detailed description of the physics is required. Some idea of this threshold may be obtained by considering unitarity violation \cite{Brivio2021,Bresciani2025A,Bresciani2025B} or from renormalization-group flow \cite{Eichhorn2012,Dupuis2021,deBrito2022}. Since we don't want to assume anything about the UV physics, we also neglect the real part of the ALP self-energy which would ordinarily appear in the denominator of Eq.~\eqref{eq:axion_crossSection} and cause the position of the resonance to be shifted by some factor.

\setcounter{equation}{0}
\renewcommand{\theequation}{B\arabic{equation}}
\subsection{B. Pseudoscalar ALP-mediated polarization and angle resolved differential cross section}
The QED BW cross section is sensitive to the polarizations of the incoming and outgoing particles as well as the scattering angle, while the ALP-mediated BW cross section depends only on the polarization. In order to compute the combined cross section, the individual amplitudes for all possible polarization and scattering angles must be calculated and combined, prior to summing over spin, averaging over photon polarization and integration over solid angle.
In principle, the angle dependent amplitude may be calculated analytically, for the s-channel ALP-mediated process and the t- and u-channel QED processes as follows. 
For the ALP-mediated process, the computation of amplitude $\mathcal{M}_{ijkl}$, for initial polarizations (i and j), and final spin projection in a fixed $z$-basis (k and l) may be reduced to calculating the product of three scalar quantities: a factor $g_{a\gamma\gamma}g_{aee}/(s-m_a^2+i\sqrt{s}\Gamma(s))$, a polarization selection rule $\epsilon^{i}_\nu \epsilon^{j}_\rho k^{(1)}_\mu k^{(2)}_\lambda \epsilon^{\mu\nu\lambda\rho}$ and the spinor bilinear form $\bar{u}^{k}(p_1)\gamma^5\nu^{l}(p_2)$.
The selection rule for $i=1$ and $j=0$,
\begin{equation}
    \epsilon^{1}_\nu \epsilon^{0}_\rho k^{(1)}_\mu k^{(2)}_\lambda \epsilon^{\mu\nu\lambda\rho} = 
\begin{vmatrix}
    \sqrt{s}/2 & 0 & \sqrt{s}/2 & 0 \\
    \sqrt{s}/2 & 0 & -\sqrt{s}/2 & 0 \\
    0 & 1 & 0 & 0 \\
    0 & 0 & 0 & 1 \\
\end{vmatrix}
= \frac{s}{2},
\end{equation}
in the COM frame and considering a linear polarization basis. For $i=0$ and $j=1$ this is $-\frac{s}{2}$ and for $i=j$ this is zero.

The spinor bilinear,
\begin{equation}
\begin{split}
\bar{u}^{k}(p_1)\gamma^5v^{l}(p_2) =& (E + m_e)
\begin{pmatrix}
    \chi^{k\dagger} \quad
    \dfrac{(\bm{\sigma} \cdot \bm{p}_1  \chi^k)^\dagger}{E + m_e}  
\end{pmatrix}
\gamma^0\gamma^5
\begin{pmatrix}
\dfrac{\bm{\sigma} \cdot \bm{p}_2}{E + m_e} \, \chi^l \\
\chi^l
\end{pmatrix} \\
=& (E + m_e)
\begin{pmatrix}
    \chi^{k\dagger} \quad
    \dfrac{(\bm{\sigma} \cdot \bm{p}_1  \chi^k)^\dagger}{E + m_e} 
\end{pmatrix}
\begin{pmatrix}
\chi^l \\
-\dfrac{\bm{\sigma} \cdot \bm{p}_2}{E + m_e} \, \chi^l
\end{pmatrix} \\
=& (E + m_e) \left(\chi^{k\dagger}\chi^l  -   \frac{(\bm{\sigma} \cdot \bm{p}_1 \chi^{k})^\dagger (\bm{\sigma} \cdot \bm{p}_2 \chi^l)}{(E+m_e)^2}   \right),
\end{split}
\end{equation}
where $E = \sqrt{s}/2$, $\bm{p}_1=-\bm{p}_2 = (p\sin(\theta)\cos(\varphi),p\sin(\theta)\sin(\varphi),p\cos(\theta))$ where $p = \sqrt{s/4-m_e^2}$ and $\theta$ and $\phi$ describe the scattering plane and angle. $\bm{\sigma}$ is the Pauli vector and $\chi$ are Pauli spinors. If we consider Pauli spinors in the z-eigenbasis for spin, $\chi^0 = (1~~0)^\mathrm{T}$ and $\chi^1 = (0~~1)^\mathrm{T}$ then,
\begin{equation}
    \bar{u}^{k}(p_1)\gamma^5 v^{l}(p_2) = \delta_{kl}\sqrt{s}
\end{equation}
We can combine these components to get,
\begin{equation}\label{EM:polxs}
    \mathcal{M}^a_{ijkl} = -\varepsilon_{ij}\delta_{kl} \left(\frac{s\sqrt{s}}{2}\right) \frac{g_{a\gamma\gamma}g_{aee}}{s-m_a^2+i\sqrt{s}\Gamma(s)}
\end{equation}
For the QED t- and u-channel processes, similar computations can be performed. However, in that case, since the interaction between the electron and photon fields isn't mediated by a pseudoscalar, the polarization combinatorics cannot be disentangled so neatly. Hence, for each channel, each of the 16 amplitudes should be calculated separately.

\setcounter{equation}{0}
\renewcommand{\theequation}{C\arabic{equation}}
\subsection{C. Pseudoscalar ALP-QED interference}
Although useful in the case where one may not want to include all combinations of spin and polarization in their calculation, the amplitudes derived in the previous section are cumbersome. Therefore, in this section, we calculate a more compact form for the full, spin summed and polarisation averaged, combined QED and ALP BW cross section. 

\begin{figure}
    \centering
    \includegraphics[width=0.99\linewidth]{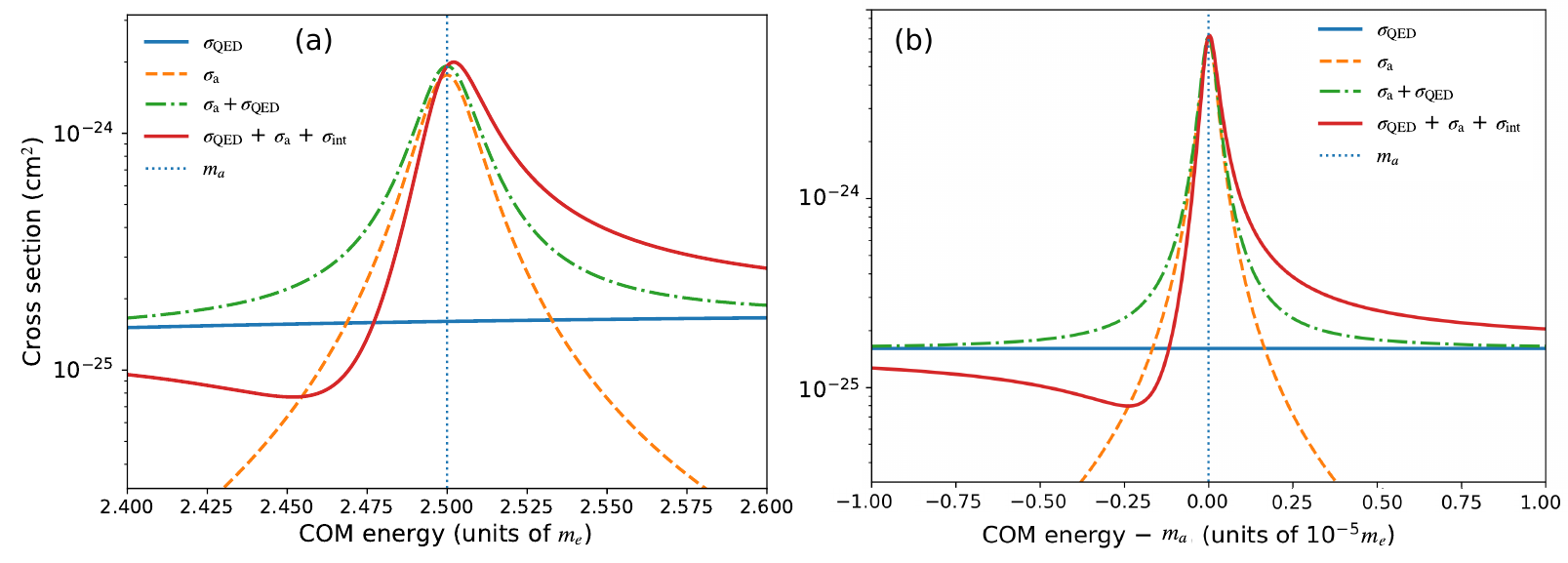}
    \caption{Different cross section contributions for the BW process governed QED alongside a pseudoscalar ALP with (a) $m_a=2.5 m_e$, $g_{aee}=1\times 10^{-2}$ and $g_{a\gamma\gamma}=10^3$ GeV$^{-1}$ as well as some less excluded parameters (b) $m_a=2.5 m_e$, $g_{aee}=1\times 10^{-4}$ and $g_{a\gamma\gamma}=5$ GeV$^{-1}$.  }
    \label{fig:interference}
\end{figure}

We start from Eq.~\eqref{eq:coherent} and expand the modulus squared of the combined amplitudes, such that
\begin{equation}\label{eq:cross_section_breakdown}
\begin{split}
    \frac{d\sigma}{d\Omega} &= \frac{\sqrt{s-4m_e^2}}{256\pi^2s\sqrt{s}} \sum_{i,j,k,l}
    \lvert \mathcal{M}^{(ijkl)}_{QED}(s,\theta,\phi)+ \mathcal{M}^{(ijkl)}_{a}(s)\rvert^2\\ &= 
    \frac{\sqrt{s-4m_e^2}}{256\pi^2s\sqrt{s}}\sum_{i,j,k,l} \left(
    \lvert \mathcal{M}^{(ijkl)}_{QED}(s,\theta,\phi)\rvert^2+ \lvert\mathcal{M}^{(ijkl)}_{a}(s)\rvert^2 + 2 \mathcal{R}\left\{\mathcal{M}^{(ijkl)}_{QED}\mathcal{M}^{(ijkl)\ast}_{a}\right\}\right).
\end{split}
\end{equation}
We already know the first two terms from Eq.~\eqref{eq:qed_crossSection} and Eq.~\eqref{eq:axion_crossSection}, respectively. It remains to calculate the interference term, which can be achieved via similar trace methods.
In particular, we need the real part of 
\begin{equation}
\begin{split}
    \sum_{i,j,k,l} \mathcal{M}^{(ijkl)}_{QED}\mathcal{M}^{(ijkl)\ast}_{a} &= 
    \sum_{i,j,k,l}\left[-\frac{ie^2 \mathcal{B}^{(t)}_{ijkl}}{t-m_e^2} - \frac{ie^2 \mathcal{B}^{(u)}_{ijkl}}{u-m_e^2}\right]  \left[\frac{g_{a\gamma\gamma}g_{aee}\epsilon^{(1)i}_\nu \epsilon^{(2)j}_\rho k^{(1)}_\mu k^{(2)}_\lambda \epsilon^{\mu\nu\lambda\rho}}{s - m_a^2 +i\sqrt{s}\Gamma(s)} \bar{u}^{k}(p_1)\gamma^5 v^{l}(p_2)\right]^\ast\\ &=\frac{-ie^2g_{a\gamma\gamma}g_{aee}}{s - m_a^2 -i\sqrt{s}\Gamma(s)}\left[\frac{1}{t-m_e^2}\sum_{i,j,k,l}\mathcal{B}^{(t)}_{ijkl}\Phi^{ijkl~\ast} +  \frac{1}{u-m_e^2}\sum_{i,j,k,l} \mathcal{B}^{(u)}_{ijkl}\Phi^{ijkl~\ast}\right],
\end{split}
\end{equation}
where 
\begin{equation}
\begin{split}
    \Phi^{ijkl~\ast} &=  [\epsilon^{(1)i}_\nu \epsilon^{(2)j}_\rho k^{(1)}_\mu k^{(2)}_\lambda \epsilon^{\mu\nu\lambda\rho} \bar{u}^{k}(p_1)\gamma^5 v^{l}(p_2)]^\ast\\
    &= -\epsilon^{(1)i\ast}_\nu \epsilon^{(2)j\ast}_\rho k^{(1)}_\mu k^{(2)}_\lambda \epsilon^{\mu\nu\lambda\rho} \bar{v}^{l}(p_2)\gamma^5 u^{k}(p_1).
\end{split}
\end{equation}
and the QED amplitude $\mathcal{M}^{(ijkl)}_{QED}$ is discussed in Supplementary Material I.
These sums over polarization may be calculated separately, giving
\begin{equation}\label{eq:polarization_sum_interference}
    \sum_{i,j,k,l} \mathcal{B}^{(t/u)}_{ijkl}\Phi^{ijkl~\ast} = -2im_es^2,
\end{equation}
(additional details are provided in Supplementary Material II) and letting us conclude that 
\begin{equation}
\sum_{i,j,k,l}\mathcal M^{(ijkl)}_{\rm QED}\mathcal M_a^{(ijkl)\ast}=\frac{-2 e^2 g_{a\gamma\gamma}g_{aee}m_e s^2}{s-m_a^2-i \sqrt s\Gamma(s)}
\left[\frac{1}{t-m_e^2}+\frac{1}{u-m_e^2}\right].  
\end{equation}
We can then use this to calculate the interference term of Eq.~\eqref{eq:cross_section_breakdown}
\begin{equation}
\begin{split}
2 \mathcal{R}\left\{\sum_{i,j,k,l}\mathcal{M}^{(ijkl)}_{QED}\mathcal{M}^{(ijkl)\ast}_{a}\right\} 
&=\frac{-4 e^2 g_{a\gamma\gamma}g_{aee}m_e s^2}{(s-m_a^2)^2+s\Gamma(s)^2}  \left[\frac{1}{t-m_e^2}+\frac{1}{u-m_e^2}\right] \mathcal{R}\left\{s-m_a^2 +i\sqrt{s}\Gamma(s)\right\}\\
&= \frac{-4 e^2 g_{a\gamma\gamma}g_{aee}m_e s^2 (s-m_a^2)}{(s-m_a^2)^2+s\Gamma(s)^2}  \left[\frac{1}{t-m_e^2}+\frac{1}{u-m_e^2}\right] .
\end{split}
\end{equation}
The interference contribution to the differential cross section is then just 
\begin{equation}
\frac{d\sigma_\mathrm{int}}{d\Omega} 
= \frac{-\sqrt{s-4m_e^2}}{64\pi^2\sqrt{s}} \frac{ e^2 g_{a\gamma\gamma}g_{aee}m_e s (s-m_a^2)}{(s-m_a^2)^2+s\Gamma(s)^2}  \left[\frac{1}{t-m_e^2}+\frac{1}{u-m_e^2}\right],
\end{equation}
which should be integrated over solid angle to find the effect of interference on the total cross section.
For simplicity, here we consider beams propagating in the $\hat{z}$ axis and scattering with momenta $\bm{p}_1=-\bm{p}_2 = (p\sin(\theta)\cos(\varphi),p\sin(\theta)\sin(\varphi),p\cos(\theta))$ where $p = \sqrt{s/4-m_e^2}$. Since
\begin{equation}
    t-m_e^2 = -\frac{s}{2}(1-\beta \cos{(\theta)}),
\end{equation}
and 
\begin{equation}
    u-m_e^2 = -\frac{s}{2}(1+\beta \cos{(\theta)}), 
\end{equation}
we can write
\begin{equation}
    \frac{1}{t-m_e^2}+\frac{1}{u-m_e^2} 
    = -\frac{4}{s} \frac{1}{1-\beta^2\cos^2{(\theta)}},
\end{equation}
where $\beta = \sqrt{1-4m_e^2/s}$. The cross section is 
\begin{equation}
    \sigma_\mathrm{int} 
    = \int \frac{d\sigma_\mathrm{int}}{d\Omega}d\Omega 
    = 2\pi \int_{-1}^1 \frac{d\sigma_\mathrm{int}}{d\Omega} d\cos{(\theta)},
\end{equation}
which may be calculated by evaluating the integral
\begin{equation}
    \int^1_{-1}\frac{1}{1-\beta^2c^2}dc 
    = \frac{2}{\beta}\tanh^{-1}(\beta)
\end{equation}
leaving us with
\begin{equation}\label{eq:int_xsec}
    \sigma_\mathrm{int} 
    = \frac{1}{4\pi} \frac{ e^2 g_{a\gamma\gamma}g_{aee}m_e  (s-m_a^2)}{(s-m_a^2)^2+s\Gamma(s)^2}\tanh^{-1}(\beta).
\end{equation}
The different contributions to the total cross section, including that of Eq.~\eqref{eq:int_xsec}, is shown in Fig.~\ref{fig:interference}.

\setcounter{equation}{0}
\renewcommand{\theequation}{D\arabic{equation}}
\subsection{D. Scalar cross sections}
If we instead consider a scalar particle with an interaction Lagrangian
\begin{equation}
    \mathcal{L}_\textrm{int}(\phi,\psi,A) =  
     - \frac{1}{4}g_{\phi\gamma\gamma} \phi F_{\mu\nu}F^{\mu\nu} - g_{\phi ee}  \phi \bar{\psi}\psi,
\end{equation}
we get a different set of Feynman rules:
\begin{itemize}
    \item Scalar-photon vertex: $ig_{\phi\gamma\gamma}\left[(k_1\cdot k_2)g^{\mu\nu} - k_2^\mu k_1^\nu \right]$
    \item Scalar-fermion vertex: $-ig_{\phi ee}$
    \item Scalar propagator: $\frac{i}{s-m_\phi^2+i\sqrt{s}\Gamma_\phi(s)}$
\end{itemize}

Using these rules, analogously to the approaches taken for the pseudoscalar ALPs in the previous appendices, one may derive a polarization averaged and spin summed cross section for the scalar-mediated BW process. This is found to be
\begin{equation}\label{eq:scalar_xsec}
    \sigma_\phi = \frac{(g_{\phi\gamma\gamma}g_{\phi ee})^2}{64\pi}\frac{ s^2\beta^3}{|s - m_\phi^2 +i\sqrt{s}\Gamma_\phi(s)|^2},
\end{equation}
while the interference contribution is
\begin{equation}\label{eq:scalar_int}
    \sigma_{\mathrm{int},\phi} = \frac{1}{4\pi} \frac{ e^2 g_{\phi\gamma\gamma}g_{\phi ee}m_e  (s-m_\phi^2)}{(s-m_\phi^2)^2+s\Gamma_\phi(s)^2}\beta^2\tanh^{-1}(\beta).
\end{equation}

\newpage
\section*{Supplementary Material}
\setcounter{equation}{0}
\renewcommand{\theequation}{$S_\mathrm{I}$\arabic{equation}}
\subsection{I. QED amplitudes}
In this section, we explain how to compute the quantum mechanical amplitude associated with the t- and u-channel polarized Breit-Wheeler process. The amplitudes may be written as 
\begin{equation}
    \mathcal{M}_\textrm{QED}^{(ijkl)} = -\frac{ie^2 \mathcal{B}^{(t)}_{ijkl}}{t-m_e^2} - \frac{ie^2 \mathcal{B}^{(u)}_{ijkl}}{u-m_e^2}
\end{equation}
These $\mathcal{B}$ are spinor bilinears
\begin{equation}
  \mathcal{B}^{(t/u)}_{ijkl} = \bar{u}_{k}(p_1)
  \epsilon_{i/j}^\mu \gamma_\mu 
  \left[q_{(t/u)}^\lambda\gamma_\lambda+m_e\right]
  \epsilon_{j/i}^\nu \gamma_\nu
  v_{l}(p_2) 
\end{equation}
which can be written in terms of $s$, $\theta$, $\phi$,  a spinor normalization factor,
\begin{equation}
    \mathcal{N}=\sqrt{m_e + \frac{\sqrt{s}}{2}},
\end{equation}
and the magnitude of the fermion momenta
\begin{equation}
    p = \sqrt{-m_e^{2} + \frac{s}{4}}.
\end{equation}
We now enumerate all possible combinations of photon polarization (0 and 1 represent orthogonal linear polarizations in the $\hat{y}$ and $\hat{z}$ direction, respectively, while the beam propagates along the $\hat{x}$ axis) and electron/positron spins (0 and 1 represent spin up and down, respectively, in the $z$ basis for electrons and positrons with momenta $\bm{p}_1=-\bm{p}_2 = (p\sin(\theta)\cos(\varphi),p\sin(\theta)\sin(\varphi),p\cos(\theta))$ where $p = \sqrt{s/4-m_e^2}$) for the bilinear and get
\begin{equation}
\mathcal{B}^{(t/u)}_{0000}=-\frac{p^{2}\sin\!\left(\theta\right)\cos\!\left(\theta\right)}{\mathcal{N}^{2}}\left[4p\sin^{2}\!\left(\phi\right)\sin\!\left(\theta\right)\pm\sqrt{s}\cos\!\left(\phi\right)\right],
\end{equation}

\begin{equation}
\mathcal{B}^{(t/u)}_{1000}=-i m_e\sqrt{s}+\frac{p\sin\!\left(\phi\right)\sin\!\left(\theta\right)}{\mathcal{N}^{2}}\left[\sqrt{s}\mathcal{N}^{2}-4p^{2}\cos^{2}\!\left(\theta\right)\right],
\end{equation}

\begin{equation}
\mathcal{B}^{(t/u)}_{0100}=i m_e\sqrt{s}+\frac{p\sin\!\left(\phi\right)\sin\!\left(\theta\right)}{\mathcal{N}^{2}}\left[\sqrt{s}\mathcal{N}^{2}-4p^{2}\cos^{2}\!\left(\theta\right)\right],
\end{equation}

\begin{equation}
\begin{split}
\mathcal{B}^{(t/u)}_{0010}=&2 i p\sqrt{s}\sin\!\left(\phi\right)\sin\!\left(\theta\right)-\frac{4 p^{3}\sin^{2}\!\left(\phi\right) \sin^{3}\!\left(\theta\right)e^{i\phi}}{\mathcal{N}^{2}}\\&\pm \sqrt{s}\left[m_e+\frac{p^{2}}{\mathcal{N}^{2}}\left(1- \cos\!\left(\phi\right)\sin^{2}\!\left(\theta\right) e^{i\phi}\right)\right],
\end{split}
\end{equation}

\begin{equation}
\begin{split}
\mathcal{B}^{(t/u)}_{0001}=&-2 i p\sqrt{s}\sin\!\left(\phi\right)\sin\!\left(\theta\right)-\frac{4 p^{3}\sin^{2}\!\left(\phi\right)\sin^{3}\!\left(\theta\right) e^{-i\phi}}{\mathcal{N}^{2}} \\&\pm \sqrt{s}\left[m_e+\frac{p^{2}}{\mathcal{N}^{2}}\left(1-\cos\!\left(\phi\right)\sin^{2}\!\left(\theta\right)e^{-i\phi}\right)\right],
\end{split}
\end{equation}

\begin{equation}
\mathcal{B}^{(t/u)}_{1100}=2p\sqrt{s}\cos\!\left(\theta\right)-\frac{p^{2}\cos\!\left(\theta\right)}{\mathcal{N}^{2}}\left[4p\cos^{2}\!\left(\theta\right)\pm \sqrt{s}\sin\!\left(\theta\right)\cos\!\left(\phi\right)\right],
\end{equation}

\begin{equation}
\mathcal{B}^{(t/u)}_{1010}=i p\sqrt{s}\cos\!\left(\theta\right)-\frac{4p^{3}\sin\!\left(\phi\right)\sin^{2}\!\left(\theta\right)\cos\!\left(\theta\right)}{\mathcal{N}^{2}}e^{i\phi},
\end{equation}

\begin{equation}
\mathcal{B}^{(t/u)}_{1001}=-i p\sqrt{s}\cos\!\left(\theta\right)-\frac{4p^{3}\sin\!\left(\phi\right)\sin^{2}\!\left(\theta\right)\cos\!\left(\theta\right)}{\mathcal{N}^{2}}e^{-i\phi},
\end{equation}

\begin{equation}
\mathcal{B}^{(t/u)}_{0110}=i p\sqrt{s}\cos\!\left(\theta\right)-\frac{4p^{3}\sin\!\left(\phi\right)\sin^{2}\!\left(\theta\right)\cos\!\left(\theta\right)}{\mathcal{N}^{2}}e^{i\phi},
\end{equation}

\begin{equation}
\mathcal{B}^{(t/u)}_{0101}=-i p\sqrt{s}\cos\!\left(\theta\right)-\frac{4p^{3}\sin\!\left(\phi\right)\sin^{2}\!\left(\theta\right)\cos\!\left(\theta\right)}{\mathcal{N}^{2}}e^{-i\phi},
\end{equation}

\begin{equation}
\mathcal{B}^{(t/u)}_{0011}=\frac{p^{2}\sin\!\left(\theta\right)\cos\!\left(\theta\right)}{\mathcal{N}^{2}}\left[4p\sin^{2}\!\left(\phi\right)\sin\!\left(\theta\right)\pm\sqrt{s}\cos\!\left(\phi\right)\right],
\end{equation}

\begin{equation}
\mathcal{B}^{(t/u)}_{1110}=\pm \frac{s}{2}-\frac{p^{2}e^{i\phi}}{\mathcal{N}^{2}}\left[4p\sin\!\left(\theta\right)\cos^{2}\!\left(\theta\right)\pm \sqrt{s}\cos\!\left(\phi\right)\sin^{2}\!\left(\theta\right)\right],
\end{equation}

\begin{equation}
\mathcal{B}^{(t/u)}_{1101}=\pm \frac{s}{2}-\frac{p^{2}e^{-i\phi}}{\mathcal{N}^{2}}\left[4p\sin\!\left(\theta\right)\cos^{2}\!\left(\theta\right)\pm \sqrt{s}\cos\!\left(\phi\right)\sin^{2}\!\left(\theta\right)\right],
\end{equation}

\begin{equation}
\mathcal{B}^{(t/u)}_{1011}=-i m_e\sqrt{s}-\frac{p\sin\!\left(\phi\right)\sin\!\left(\theta\right)}{\mathcal{N}^{2}}\left[\sqrt{s}\mathcal{N}^{2}-4p^{2}\cos^{2}\!\left(\theta\right)\right],
\end{equation}

\begin{equation}
\mathcal{B}^{(t/u)}_{0111}=i m_e\sqrt{s}-\frac{p\sin\!\left(\phi\right)\sin\!\left(\theta\right)}{\mathcal{N}^{2}}\left[\sqrt{s}\mathcal{N}^{2}-4p^{2}\cos^{2}\!\left(\theta\right)\right],
\end{equation}

\begin{equation}
\mathcal{B}^{(t/u)}_{1111}=-2p\sqrt{s}\cos\!\left(\theta\right)+\frac{p^{2}\cos\!\left(\theta\right)}{\mathcal{N}^{2}}\left[4p\cos^{2}\!\left(\theta\right)\pm \sqrt{s}\sin\!\left(\theta\right)\cos\!\left(\phi\right)\right].
\end{equation}
Where multiple signs were listed, the upper (lower) sign represents the sign associated with the t-(u-)channel process.

These spinor bilinears have some polarization structure. In particular,
\FloatBarrier
\begin{figure}[h]
\centering
\begin{tikzpicture}[node distance=0.8cm and 1.8cm]
\tikzset{
    Bnode/.style={
        draw,
        rounded corners,
        minimum width=3.0cm,
        minimum height=0.85cm,
        align=center
    },
    rel/.style={-{Latex[length=2mm]}, thick}
}
\node[Bnode] (B1000) {$\mathcal{B}^{(t/u)}_{1000}$};
\node[Bnode, right=of B1000] (B0100) {$\mathcal{B}^{(t/u)}_{0100}$};
\node[Bnode, below=of B1000] (B0111) {$\mathcal{B}^{(t/u)}_{0111}$};
\node[Bnode, below=of B0100] (B1011) {$\mathcal{B}^{(t/u)}_{1011}$};

\draw[rel] (B1000) -- node[above] {$*$} (B0100);
\draw[rel] (B1000) -- node[left] {$-$} (B0111);
\draw[rel] (B0100) -- node[right] {$-$} (B1011);
\draw[rel] (B0111) -- node[below] {$*$} (B1011);

\begin{scope}[xshift=9.5cm]
\node[Bnode] (B1010) {$\mathcal{B}^{(t/u)}_{1010}$};
\node[Bnode, right=of B1010] (B0110) {$\mathcal{B}^{(t/u)}_{0110}$};
\node[Bnode, below=of B1010] (B1001) {$\mathcal{B}^{(t/u)}_{1001}$};
\node[Bnode, below=of B0110] (B0101) {$\mathcal{B}^{(t/u)}_{0101}$};

\draw[rel] (B1010) -- node[above] {$=$} (B0110);
\draw[rel] (B1001) -- node[below] {$=$} (B0101);
\draw[rel] (B1010) -- node[left] {$*$} (B1001);
\draw[rel] (B0110) -- node[right] {$*$} (B0101);
\end{scope}
\end{tikzpicture}
\end{figure}
\begin{figure}[h]
\centering
\begin{tikzpicture}[node distance=0.8cm and 1.8cm]
\tikzset{
    Bnode/.style={
        draw,
        rounded corners,
        minimum width=3.0cm,
        minimum height=0.85cm,
        align=center
    },
    rel/.style={-{Latex[length=2mm]}, thick}
}
\node[Bnode] (B0000) {$\mathcal{B}^{(t/u)}_{0000}$};
\node[Bnode, right=of B0000] (B0011) {$\mathcal{B}^{(t/u)}_{0011}$};
\draw[rel] (B0000) -- node[above] {$-$} (B0011);
\begin{scope}[xshift=9.5cm]
\node[Bnode] (B1100) {$\mathcal{B}^{(t/u)}_{1100}$};
\node[Bnode, right=of B1100] (B1111) {$\mathcal{B}^{(t/u)}_{1111}$};
\draw[rel] (B1100) -- node[above] {$-$} (B1111);
\end{scope}
\begin{scope}[yshift=-1.5cm]
\node[Bnode] (B0010) {$\mathcal{B}^{(t/u)}_{0010}$};
\node[Bnode, right=of B0010] (B0001) {$\mathcal{B}^{(t/u)}_{0001}$};
\draw[rel] (B0010) -- node[above] {$*$} (B0001);
\end{scope}
\begin{scope}[xshift=9.5cm,yshift=-1.5cm]
\node[Bnode] (B1110) {$\mathcal{B}^{(t/u)}_{1110}$};
\node[Bnode, right=of B1110] (B1101) {$\mathcal{B}^{(t/u)}_{1101}$};
\draw[rel] (B1110) -- node[above] {$*$} (B1101);
\end{scope}
\end{tikzpicture}
\end{figure}
\FloatBarrier
\noindent where the $\ast$ ($-$) [$=$] denotes that the quantities are related by complex conjugation (multiplication by negative one) [equality].

\newpage
\setcounter{equation}{0}
\renewcommand{\theequation}{$S_\mathrm{II}$\arabic{equation}}
\subsection{II. Spin and polarisation summations}
We want to evaluate the polarisation sum in Eq.~\eqref{eq:polarization_sum_interference}.
The external photon polarisation sums are
\begin{equation}
\sum_i \epsilon^{(1)i}_\alpha \epsilon^{(1)i,\ast}_\nu=-g_{\alpha\nu},
\qquad
\sum_j \epsilon^{(2)j}_\beta \epsilon^{(2)j,\ast}_\rho=-g_{\beta\rho},
\end{equation}
and the spin sums are
\begin{equation}
\sum_k u^k(p_1)\bar u^k(p_1)=p_{1\alpha}\gamma^\alpha + m_e,
\qquad
\sum_l v^l(p_2)\bar v^l(p_2)=p_{2\alpha}\gamma^\alpha - m_e.
\end{equation}

For the $t$-channel, we write $q_t = p_1-k^{(1)}$ and then
\begin{equation}
\mathcal{B}^{(t)}_{ijkl}=
\bar{u}^{k}(p_1)\epsilon^{(1)i\alpha}\gamma_\alpha\left[q_t^\sigma\gamma_\sigma+ m_e\right]\epsilon^{(2)j\beta}\gamma_\beta v^{l}(p_2),
\end{equation}
while
\begin{equation}
\begin{split}
\Phi^{ijkl,\ast}
=-\epsilon^{(1)i,\ast}_\nu\epsilon^{(2)j,\ast}_\rho k^{(1)}_\mu k^{(2)}_\lambda \epsilon^{\mu\nu\lambda\rho} \bar{v}^{l}(p_2)\gamma^5 u^{k}(p_1).
\end{split}
\end{equation}
In this form, our polarisation sum is 
\begin{equation}
\begin{split}
\sum_{i,j,k,l}\mathcal{B}^{(t)}_{ijkl}\Phi^{ijkl,\ast}
&=-\epsilon^{\mu\nu\lambda\rho}k^{(1)}_\mu k^{(2)}_\lambda \sum_{k,l}\bar u^k(p_1)\gamma_\nu\left[q_t^\sigma\gamma_\sigma + m_e\right]\gamma_\rho v^l(p_2)\bar v^l(p_2)\gamma^5 u^k(p_1)\\
&=-\epsilon^{\mu\nu\lambda\rho}k^{(1)}_\mu k^{(2)}_\lambda T^{(t)}_{\nu\rho},
\end{split}
\end{equation}
where
\begin{equation}
T^{(t)}_{\nu\rho}
=\mathrm{Tr}\left[\left(p_{1\alpha}\gamma^\alpha+m_e\right)\gamma_\nu\left(q_t^\sigma\gamma_\sigma+m_e\right)\gamma_\rho\left(p_{2\beta}\gamma^\beta-m_e\right)\gamma^5\right].
\end{equation}
Only the terms containing exactly four ordinary gamma matrices in addition to $\gamma^5$ are non-zero. Hence
\begin{equation}
\begin{split}
T^{(t)}_{\nu\rho}
&=m_e q_t^a p_2^b\mathrm{Tr}\left[\gamma_\nu\gamma_a\gamma_\rho\gamma_b\gamma^5\right]+m_e p_1^a p_2^b\mathrm{Tr}\left[\gamma_a\gamma_\nu\gamma_\rho\gamma_b\gamma^5\right]-m_e p_1^a q_t^b\mathrm{Tr}\left[\gamma_a\gamma_\nu\gamma_b\gamma_\rho\gamma^5\right].
\end{split}
\end{equation}
Using
\begin{equation}
\mathrm{Tr}\left[\gamma^\alpha\gamma^\beta\gamma^\gamma\gamma^\delta\gamma^5\right]=-4i\epsilon^{\alpha\beta\gamma\delta},
\end{equation}
this becomes
\begin{equation}
T^{(t)}_{\nu\rho}
=-4im_e\epsilon_{a\nu b\rho}\left[q_t^a p_2^b-p_1^a p_2^b-p_1^a q_t^b\right].
\end{equation}
Now, inserting $q_t=p_1-k^{(1)}$ gives
\begin{equation}
\begin{split}
q_t^a p_2^b-p_1^a p_2^b-p_1^a q_t^b
=\left(p_1^a-k^{(1)a}\right)p_2^b-p_1^a p_2^b-p_1^a\left(p_1^b-k^{(1)b}\right)
=-k^{(1)a}p_2^b-p_1^a p_1^b+p_1^a k^{(1)b}.
\end{split}
\end{equation}
The term proportional to $p_1^a p_1^b$ vanishes after contraction with $\epsilon_{a\nu b\rho}$, because it is symmetric under $a\leftrightarrow b$.
Using momentum conservation,
\begin{equation}
p_1+p_2=k^{(1)}+k^{(2)},
\qquad
p_2=k^{(1)}+k^{(2)}-p_1,
\end{equation}
we find
\begin{equation}
\begin{split}
-k^{(1)a}p_2^b+p_1^a k^{(1)b}
=-k^{(1)a}\left(k^{(1)b}+k^{(2)b}-p_1^b\right)+p_1^a k^{(1)b}
=k^{(1)a}k^{(1)b}-k^{(1)a}k^{(2)b}+k^{(1)a}p_1^b+p_1^a k^{(1)b}.
\end{split}
\end{equation}
The terms proportional to $k^{(1)a}k^{(1)b}$ and $k^{(1)a}p_1^b+p_1^a k^{(1)b}$ are symmetric in $a,b$, and therefore vanish when contracted with $\epsilon_{a\nu b\rho}$. 
Hence, the only surviving contribution is
\begin{equation}
q_t^a p_2^b-p_1^a p_2^b-p_1^a q_t^b
=-k^{(1)a}k^{(2)b}.
\end{equation}
Therefore
\begin{equation}
T^{(t)}_{\nu\rho}
=4im_e\epsilon_{a\nu b\rho}k^{(1)a}k^{(2)b}.
\end{equation}
Substituting this back into the spin and polarisation sum gives
\begin{equation}
\begin{split}
\sum_{i,j,k,l}\mathcal{B}^{(t)}_{ijkl}\Phi^{ijkl,\ast}
=-\epsilon^{\mu\nu\lambda\rho}k^{(1)}_\mu k^{(2)}_\lambda\left[4im_e\epsilon_{a\nu b\rho}k^{(1)a}k^{(2)b}\right]
=-4im_e\epsilon^{\mu\nu\lambda\rho}\epsilon_{a\nu b\rho}k^{(1)}_\mu k^{(2)}_\lambda k^{(1)a} k^{(2)b}.
\end{split}
\end{equation}
The remaining Levi-Civita contraction is
\begin{equation}
\epsilon^{\mu\nu\lambda\rho} \epsilon_{a\nu b\rho}k^{(1)}_\mu k^{(2)}_\lambda k^{(1)a} k^{(2)b}
=\frac{s^2}{2}.
\end{equation}
This follows from
\begin{equation}
k^{(1)2}=k^{(2)2}=0,
\qquad
2k^{(1)}\cdot k^{(2)}=s.
\end{equation}
Thus
\begin{equation}
\sum_{i,j,k,l}\mathcal{B}^{(t)}_{ijkl}\Phi^{ijkl,\ast}
=-2im_e s^2.
\end{equation}

The $u$-channel is similar; we instead define $q_u=p_1-k^{(2)}$ and write
\begin{equation}
\mathcal{B}^{(u)}_{ijkl}
=\bar{u}^{k}(p_1)\epsilon^{(2)j\alpha}\gamma_\alpha\left[q_u^\sigma\gamma_\sigma + m_e\right]\epsilon^{(1)i\beta}\gamma_\beta v^{l}(p_2).
\end{equation}
Repeating the same steps gives
\begin{equation}
\begin{split}
\sum_{i,j,k,l}\mathcal{B}^{(u)}_{ijkl}\Phi^{ijkl,\ast}
&=-\epsilon^{\mu\nu\lambda\rho}k^{(1)}_\mu k^{(2)}_\lambda T^{(u)}_{\rho\nu},
\end{split}
\end{equation}
where
\begin{equation}
T^{(u)}_{\rho\nu}
=\mathrm{Tr}\left[\left(p_{1\alpha}\gamma^\alpha+m_e\right)\gamma_\rho\left(q_u^\sigma\gamma_\sigma+m_e\right)\gamma_\nu\left(p_{2\beta}\gamma^\beta-m_e\right)\gamma^5\right].
\end{equation}
The result is the same expression with the two photon labels exchanged:
\begin{equation}
T^{(u)}_{\rho\nu}
=4im_e\epsilon_{a\rho b\nu}k^{(2)a}k^{(1)b}.
\end{equation}
Hence
\begin{equation}
\begin{split}
\sum_{i,j,k,l}\mathcal{B}^{(u)}_{ijkl}\Phi^{ijkl,\ast}
&=-4im_e\epsilon^{\mu\nu\lambda\rho}\epsilon_{a\rho b\nu}k^{(1)}_\mu k^{(2)}_\lambda k^{(2)a}k^{(1)b}.
\end{split}
\end{equation}
Using the antisymmetry of the Levi-Civita tensor, this contraction is identical to the $t$-channel contraction. Therefore
\begin{equation}
\sum_{i,j,k,l}\mathcal{B}^{(u)}_{ijkl}\Phi^{ijkl,\ast}
=-2im_e s^2.
\end{equation}

\end{document}